\documentclass[10pt,journal,compsoc]{IEEEtran} % TMC 期刊模板

\usepackage{bm}
\usepackage{amsmath}
\usepackage{epsf}
\usepackage{graphics}
\usepackage{ amssymb }
\usepackage[dvips]{graphicx}
\usepackage{mathtools}% Loads amsmath
\usepackage{epsfig}
\usepackage{cite}
\usepackage[linesnumbered,ruled,vlined]{algorithm2e}
\usepackage{colortbl}
\usepackage{color}
\usepackage{hyperref} % 用于插入网页超链接
\usepackage{enumitem}
\usepackage{soul,xcolor}

\usepackage{bm}
\usepackage{amsmath}
\usepackage{epsf}
\usepackage{graphics}
\usepackage{ amssymb }
\usepackage[dvips]{graphicx}
\usepackage{epsfig}
\usepackage{cite}
\usepackage[linesnumbered,ruled,vlined]{algorithm2e}
\usepackage{graphicx}
\usepackage{epsfig}
\usepackage{latexsym}
\usepackage{amsfonts}
\usepackage{here}
\usepackage{rawfonts}
\usepackage[utf8]{inputenc}
\usepackage[english]{babel}
\usepackage{amsmath}
\usepackage{amsfonts}
\usepackage{amssymb}
\usepackage{color}
\usepackage{bm}
\usepackage{listings}
\usepackage{caption}
\usepackage{multicol}
\usepackage{tabularx,booktabs}
\newcolumntype{Y}{>{\centering\arraybackslash}X}
\makeatletter
\newcommand\notsotiny{\@setfontsize\notsotiny{6.31415}{7.1828}}
\makeatother
\usepackage{amssymb}
\usepackage{amsthm}
\usepackage{graphicx}
\usepackage{epstopdf}
\usepackage{listings}
\usepackage{float}
\usepackage{amsmath}
\usepackage{amssymb}
\usepackage{amsfonts}
\usepackage{epstopdf}

\usepackage{multirow}
\usepackage{amscd}
\usepackage{mathrsfs}
\usepackage{graphicx}
\usepackage{color}
\usepackage{url}
\usepackage{bm}
\usepackage{setspace}
\usepackage{footnote}
\usepackage{xcolor}
\definecolor{lightblue}{RGB}{173, 216, 230}  % 定义淡蓝色
\newtheorem{definition}{Definition}

\newtheorem{remark}{Remark}

\usepackage[]{algorithm2e}

\addto\captionsenglish{}
\usepackage{bbm}

\usepackage{multicol}
\usepackage{mathtools}

\usepackage{lipsum,graphicx,subcaption}

\usepackage{diagbox}
\usepackage{hyperref}
\usepackage[protrusion=true,expansion=true]{microtype}
\pdfoutput=1
\usepackage[font=footnotesize]{caption}
\allowdisplaybreaks[4]

\let\emptyset\varnothing

\usepackage{graphicx}
\usepackage{grffile}
\usepackage{tabularx}
\newcounter{term}[section]

\renewcommand\theterm{\alph{term}}
\makeatletter
\newcommand{\vast}{\bBigg@{4}}

\newcommand{\Vast}{\bBigg@{5}}
\makeatother
\newcommand\semiHuge{\fontsize{22.7}{31.38}\selectfont}
\usepackage{soul,xcolor}

\usepackage[most]{tcolorbox}
\tcbset{colback=yellow!10!white, colframe=black!50!black, 
        highlight math style= {enhanced, %<-- needed for the ’remember’ options
            colframe=black,colback=red!10!white,boxsep=0pt}
        }
        \allowdisplaybreaks

\begin{document} 
\title{{\semiHuge Unleashing the Power of Tree-of-Thoughts for Edge-Enabled AIGC Service Provisioning}}
\author{Zhang Liu, Shanhao Zhan, Shaowei Shen, Lianfen Huang, Qiao Xiang,\\
Ying-Jun Angela Zhang,~\IEEEmembership{Fellow,~IEEE}, and Dusit Niyato,~\IEEEmembership{Fellow,~IEEE}
\thanks{ Z. Liu (zhangliu@xmu.edu.cn) and Q. Xiang (xiangq27@gmail.com) are with the Department of Computer Science and Technology, Xiamen University, China.\\
S. Zhan (shanhao@stu.xmu.edu.cn) and S. Shen (shenshaowei@stu.xmu.edu.cn) are with the Department of Informatics and Communication Engineering, Xiamen University, China. \\
L. Huang (lfhuang@xmu.edu.cn) is with the Key Laboratory of Intelligent Manufacturing Equipment and Industrial Internet Technology, School of Information Science and Technology, Xiamen University Tan Kah Kee College, China, and also with the Department of Informatics and Communication Engineering, Xiamen University, China. \\
Y. Zhang (yjzhang@ie.cuhk.edu.hk) is with the Department of Information Engineering, The Chinese University of Hong Kong, Hong Kong. \\
D. Niyato (dniyato@ntu.edu.sg) is with the College of Computing and Data Science, Nanyang Technological University, Singapore. \\
(Corresponding author: Lianfen Huang.)} }
\maketitle
% \vspace{-9mm}
\setulcolor{red}
\setul{red}{2pt}
\setstcolor{red}   

\begin{abstract}
Delivering AI-generated content (AIGC) services fundamentally relies on the reasoning capabilities of generative AI (GenAI) models. Chain-of-Thought (CoT) enhances such reasoning by guiding models through intermediate steps, while Tree-of-Thoughts (ToT) further extends CoT by exploring multiple candidate reasoning paths simultaneously, thereby greatly improving AIGC service quality. However, generating diverse reasoning paths requires separate calls to computationally intensive GenAI models, posing significant challenges for resource-constrained user devices. In this paper, we investigate mobile edge computing-enabled AIGC service provisioning with ToT prompting. Specifically, using creative writing AIGC tasks as a case study, we first characterize the number of output tokens as a measure of computational resources in GenAI models and establish its relationship with generation delay and quality through experiments with Qwen 2.5-7B-Instruct. Afterward, we introduce a directed acyclic graph (DAG) model to accurately characterize the reasoning process of ToT prompting, where each vertex represents a thought and each directed edge denotes a transition between consecutive thoughts. We then formulate a DAG-based thought assignment problem aimed at minimizing generation delay subject to a user-adjustable quality constraint. To address this problem, we propose a diffusion-based soft actor-critic (DSAC) algorithm that innovatively integrates diffusion models to determine optimal thought assignment decisions. Through extensive simulations, we demonstrate that the proposed DSAC achieves total generation delay reductions of up to 8.32\% over PPO, 11.57\% over SAC, and 36.09\% over DDQN across various simulation settings, while reducing latency by over 80\% compared to the fully local generation baseline even under stringent quality requirements.

% The implementation of our proposed method is available at: \url{https://github.com/ZhangLiu/ToTAIGC}.

% Given that AIGC services rely on large generative AI (GenAI) models, low-altitude uncrewed aerial vehicle (UAV)-assisted mobile edge computing (MEC) has emerged as a new paradigm for offering seamless and powerful computational services. Numerical results demonstrate that the BAI-MCTS algorithm achieves a convergence rate approximately 50.44\% faster than the state-of-the-art algorithms when reaching 98\% of the optimal value. Notably, the convergence rate of the LLM-BAI-MCTS algorithm increases by over 63.32\% in dense networks.

\end{abstract}
% \vspace{-.1mm}
\vspace{-3mm}
\section{Introduction} \label{sec:intro}
\subsection{Background and Overview} \label{subsec:background}
\vspace{-.15mm}

\begin{figure}[t!]
\includegraphics[width=.48\textwidth]{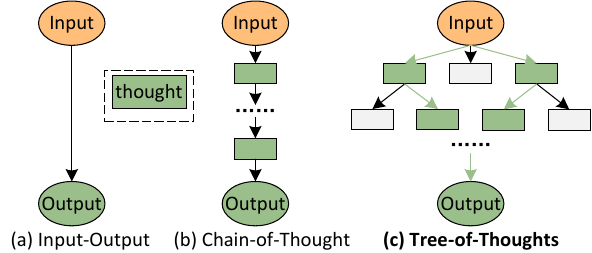}
\centering
\vspace{-1.5mm}
\caption{A schematic illustration of different prompting approaches for GenAI model inference (e.g., GPT-5.4), where each rectangular box denotes a thought—a coherent language sequence that acts as an intermediate step toward AIGC service provisioning.}
\label{fig:ToT}
% \vspace{-1mm}
\end{figure}

Driven by the rapid advancements in generative artificial intelligence (GenAI), AI-generated content (AIGC) has revolutionized the production of diverse, high-quality content~\cite{du2023exploring}. For example, ChatGPT~\cite{van2023chatgpt} can generate large volumes of content--including text, images, and videos--based on human instructions. Recently, building on the concept of \emph{Chain-of-Thought}~\cite{wei2022chain}, a new framework for GenAI model inference, \emph{Tree-of-Thoughts} (ToT)~\cite{yao2023tree}, has further enhanced the AIGC service-provisioning capabilities of GenAI models. Specifically, as shown in Fig.~\ref{fig:ToT}, unlike (a) wrapping the input in human instructions to produce the final answer directly or (b) generating a sequence of intermediate thoughts that connect the input to the final answer, (c) ToT enables multiple reasoning paths to determine the next course of action. This mechanism allows the exploration of coherent text units (i.e., thoughts) as intermediate steps toward delivering AIGC services (see Sec.~\ref{subsec:ToT} for details).

Despite the remarkable progress in AIGC and ToT, their real-world implementation continues to face numerous challenges. First, delivering AIGC services depends on the inference process of GenAI models, whose growing size and complexity pose significant challenges for deployment~\cite{liu2025two}. For instance, built on GPT-5 with approximately 52 trillion parameters, ChatGPT requires clusters of 128×80 GB A100 GPUs to generate contextually relevant responses.\footnote{\href{https://openai.com/index/introducing-gpt-5-4/}{https://openai.com/index/introducing-gpt-5-4/}} Second, ToT is resource-intensive, as each thought requires a new GenAI model call, and exploring multiple reasoning paths rapidly multiplies the number of generations. Consequently, ToT demands significantly more computation and time.

To address this, mobile cloud computing~\cite{dinh2013survey}, with its high processing speed and sustainable energy supply, enables users to access cloud-based AIGC services through the core network. However, with the massive surge in AIGC service demand--for example, ChatGPT reached 100 million active users within just two months of its launch--the long latency of cloud computing will exceed what networks can affordably support. To this end, mobile edge computing (MEC)~\cite{mao2017survey} extends cloud capabilities to the network edge by deploying computing resources close to end devices, thereby reducing service latency and alleviating the computational burden on resource-constrained users. In a typical MEC system, user devices offload computation-intensive tasks to nearby edge servers via wireless links, and the servers process the tasks and return the results to the users.

\vspace{-3mm}
\subsection{Motivation and Main Challenges} \label{subsec:challenges}
\vspace{-.15mm}
Even though edge-enabled AIGC service provisioning hold great potential, several challenges remain:

\begin{itemize}[leftmargin=4mm]
\item \textbf{\textit{Defining the notion of computational resources within GenAI models and quantifying their relationship to AIGC generation delay and quality remains challenging:}} In contrast to most MEC studies~\cite{wang2021dependent,yan2020offloading, nguyen2023dependency, zhou2023dag}, where computational resources are clearly defined as CPU cycles per second, the concept of computational resources for GenAI models remains ambiguous. It is essential to identify which controllable parameters that influence the generation time and quality of AIGC outputs. Without a well-defined expression linking GenAI computational resources to both generation delay and output quality, quantifying AIGC service provisioning becomes impossible. To address this, we focus on a specific type of AIGC--creative writing--and define the computational resources of GenAI models in terms of the number of output tokens. To model the mathematical relationships among output token count, writing generation delay, and the quality of the generated text, we conduct extensive experiments with Qwen 2.5-7B-Instruct~\cite{qwen2.5}, an open-source language model with outstanding performance, deployed locally. We then apply parameter fitting methods to develop one of the first such mathematical functions.

\item \textbf{\textit{Modeling the ToT-empowered AIGC process is challenging:}} ToT expands AIGC generation into multiple intermediate reasoning paths and requires iterative generation and evaluation of candidate thoughts, making the prompting process difficult to characterize tractably. Moreover, the reasoning structure of ToT is inherently sequential and branching, as different thoughts may lead to different subsequent paths and final outputs. To address these challenges, we introduce a directed acyclic graph (DAG) model~\cite{topcuoglu2002performance} to characterize the reasoning process of ToT prompting, where each vertex represents a thought generated by the GenAI model and each directed edge denotes the transition from one thought to the next. With this framework, we formulate the ToT-empowered AIGC process as a DAG-based thought assignment problem. The objective is to determine the optimal mapping of thoughts to edge servers, such as base stations and user devices with surplus computing resources, thereby enabling systematic optimization for low-latency, high-quality AIGC service delivery.

\item \textbf{\textit{Conventional optimization methods are ill-suited to the highly dynamic nature of edge environments:}} Due to time-varying wireless conditions, fluctuating resource availability, and changing service demands, the system state may evolve rapidly over time. Traditional heuristic methods~\cite{nguyen2023dependency,zhang2015energy} and convex relaxation approaches~\cite{mahmoodi2016optimal,guo2016energy} often rely heavily on expert knowledge and require frequent redesign or retuning to adapt to changing environments, which is both time-consuming and impractical. Recently, deep reinforcement learning (DRL) has emerged as a promising solution. However, for discrete action problems such as the DAG-based thought assignment problem, value-based DRL methods typically update the Q-network based on one-step rewards, which directly affects only individual state-action pairs and may result in slow convergence~\cite{mnih2016asynchronous}. To address this issue, we propose replacing conventional DNNs with a diffusion model~\cite{ho2020denoising}, which captures the global structure of action distributions through iterative denoising. This design enables exploration over a broader range of action candidates, thereby improving convergence speed and policy robustness.

\end{itemize} 

\vspace{-3mm}
\subsection{Summary of Contributions} \label{subsec:contributions}
\vspace{-.15mm}

\emph{To the best of our knowledge, this is the first work to investigate the modeling and optimization of ToT-prompted AIGC service provisioning in edge-enabled environments.} Our main contributions are as follows:

\begin{itemize}[leftmargin=4mm]
\item \textbf{\textit{Modeling:}} Focusing on creative writing AIGC services, we conduct extensive experiments with Qwen 2.5-7B-Instruct~\cite{qwen2.5} to develop one of the first mathematical functions linking output token count, writing generation delay, and the quality of the generated text. Building on this analytical framework, we introduce a DAG to accurately capture the reasoning process of ToT prompting and formulate a DAG-based thought assignment problem, an integer nonlinear programming (INLP) problem known to be non-convex.

\item \textbf{\textit{Solution:}} To address this problem effectively, we propose a diffusion-based soft actor-critic (DSAC) algorithm. Specifically, the DSAC algorithm leverages diffusion models--originally developed for image generation--to generate optimal thought assignment decisions. The denoising process in diffusion models enables more thorough exploration of the solution space, faster convergence, and avoidance of local optima under the heterogeneous and dynamic conditions of MEC systems.

\item \textbf{\textit{Validation:}} We evaluate the effectiveness of the proposed DSAC algorithm through extensive experiments under various simulation settings, comparing its performance against three benchmark DRL algorithms. The results demonstrate that DSAC achieves total generation delay reductions of up to 8.32\% over PPO, 11.57\% over SAC, and 36.09\% over DDQN across various simulation settings, while reducing latency by over 80\% compared to the fully local generation baseline even under stringent quality requirements, all with only a modest increase in computational overhead.

% , demonstrating that our method not only achieves a higher communication rate but also satisfies the long-term energy constraint for sustainable UAV operation.
\end{itemize}

\vspace{-3mm}
\subsection{Paper Organization} \label{subsec:organization}
\vspace{-.15mm}
The rest of the paper is organized as follows. Sec.~\ref{sec:relatedwork} reviews related work. Sec.~\ref{sec:systemmodel} presents the system model and formulates the DAG-based thought assignment optimization problem. Sec.~\ref{sec:diffusionmodel} introduces the preliminaries of diffusion models. Sec.~\ref{sec:DSAC} describes the proposed DSAC algorithm. Sec.~\ref{sec:simulation} reports the simulation results, and Sec.~\ref{sec:conclusion} concludes the paper with directions for future research.

\vspace{-3mm}
\section{Related Work} \label{sec:relatedwork}
\vspace{-.15mm}
Henceforth, we summarize the contributions of related works and highlight the aspects they have not addressed, which serve as the primary motivations for this work.

\vspace{-3mm}
\subsection{DAG Task Scheduling in MEC Networks}
\vspace{-.15mm}

The ToT prompting process unfolds through multiple interconnected thoughts and can be naturally represented as a DAG, where each vertex corresponds to a thought generated by the GenAI model and each directed edge denotes the logical progression from one thought to the next. Consequently, DAG task scheduling has been widely studied in MEC networks. For example, the authors in~\cite{liu2023rfid} investigated DAG task offloading in vehicular clouds to minimize the overall task completion time while ensuring a high execution success rate. In~\cite{huang2023joint}, the authors examined joint DAG task scheduling in UAV-enabled aerial edge computing with the objective of optimizing both the makespan and the energy consumption of DAG task execution. The authors in~\cite{9774945} studied a multiobjective DAG task scheduling problem in MEC-aided 6G networks to achieve low latency under limited energy resources. In~\cite{11271689}, the authors formulated an online multi-DAG task scheduling problem to minimize task completion time while guaranteeing server load balance. The authors in~\cite{10227271} investigated scheduling and offloading schemes for DAG tasks in MEC scenarios with the aim of minimizing completion time.

However, the execution cost of a conventional DAG task can be clearly quantified by CPU cycles, and optimization objectives--such as minimizing latency or energy consumption--are well established, supported by mature mathematical formulations. In contrast, in the ToT-empowered AIGC services, the notion of computational resources is not clearly defined, and the relationship among computational resources, AIGC generation delay, and quality lacks well-validated closed-form expressions.

\vspace{-3mm}
\subsection{Usage of Deep Reinforcement Learning in Optimization}\label{subsec:DRLshortcomings}
\vspace{-.15mm}

In recent years, learning-based algorithms, particularly DRL, have been extensively employed to enhance real-time decision-making and address complex optimization problems. The authors in~\cite{yan2020offloading} proposed a DRL framework based on an actor–critic learning structure to jointly optimize offloading decisions and resource allocation. The authors in~\cite{10497174} proposed a graph neural network-augmented DRL scheme for timely DAG task scheduling in dynamic vehicular cloud environments. The authors in~\cite{tang2020deep} integrated long short-term memory with a dueling deep Q-network to determine offloading decisions for each device, including whether to offload and, if so, which edge node to offload its task to. The authors in~\cite{wang2025joint} proposed an improved PPO algorithm to jointly optimize task-serving decisions (offloading and migration) in order to reduce latency for all mobile users in UAV-assisted MEC networks. The authors in~\cite{huang2025two} proposed a two-time-scale DRL learning approach that optimizes caching decisions at a large time-scale agent, while focusing on offloading and resource allocation at a short time-scale agent.
 
However, current DRL methods, including value-based approaches (e.g., DQN) and policy-based approaches (e.g., PPO), each have their shortcomings. Specifically, value-based DRL derives a deterministic policy by updating the deep Q-network based on one-step rewards. However, this one-step update rule directly influences only the value of the state–action pair that produced the reward, while the values of other pairs are affected only indirectly through the updated Q-network. As a result, the learning process can be slow~\cite{mnih2016asynchronous}. In contrast to value-based methods, policy-based DRL directly parameterizes the policy with a DNN and updates it through gradient ascent on the total return, enabling faster convergence. However, when tackling complex and dynamic optimization problems, the use of Monte Carlo estimation can introduce high variance, slowing down learning. In addition, conventional policy gradient methods are often data-inefficient and prone to getting stuck in local optima~\cite{doerr2019trajectory}.

\begin{figure}[t!]
\includegraphics[width=.31\textwidth]{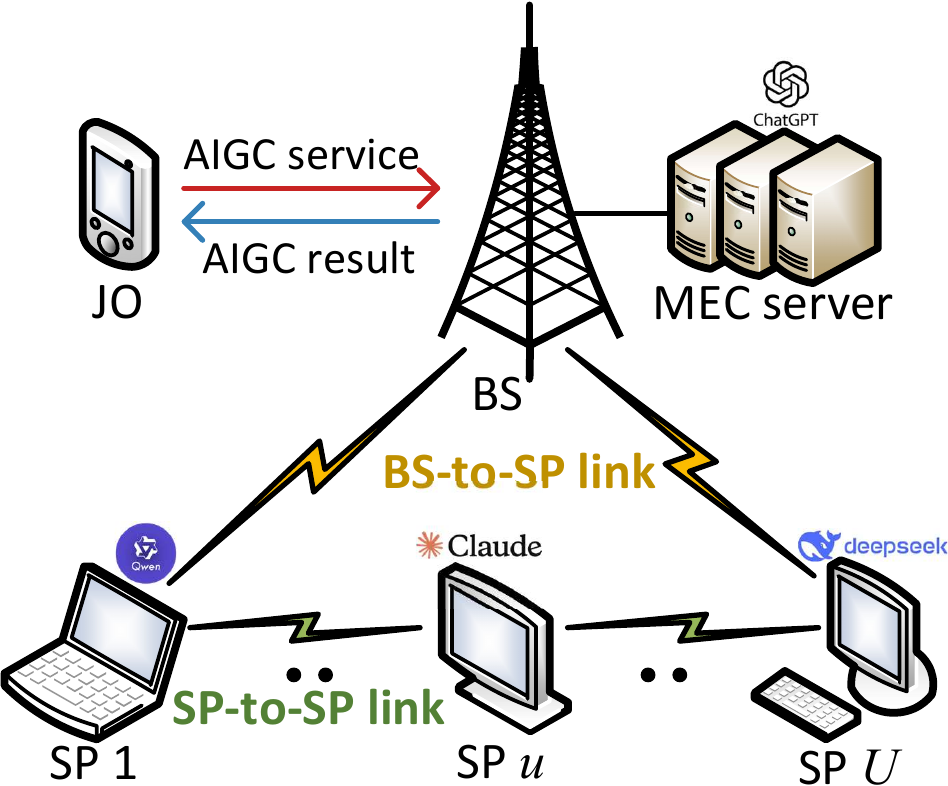}
\centering
\vspace{-1.5mm}
\caption{An illustration of edge-enabled AIGC service provisioning. The JO submits an AIGC task to the BS, which is equipped with an MEC server running a large-scale GenAI model (e.g., ChatGPT), and receives the final result via a downlink. To accelerate ToT-based AIGC delivery, the BS offloads individual thought generation tasks to SPs, each equipped with a lightweight GenAI model (e.g., Qwen, Claude, DeepSeek), via BS-to-SP and SP-to-SP links.}
\label{fig:MEC}
% \vspace{-1mm}
\end{figure}

\begin{figure*}[t!]
\vspace{-2mm}
\includegraphics[width=.98\textwidth]{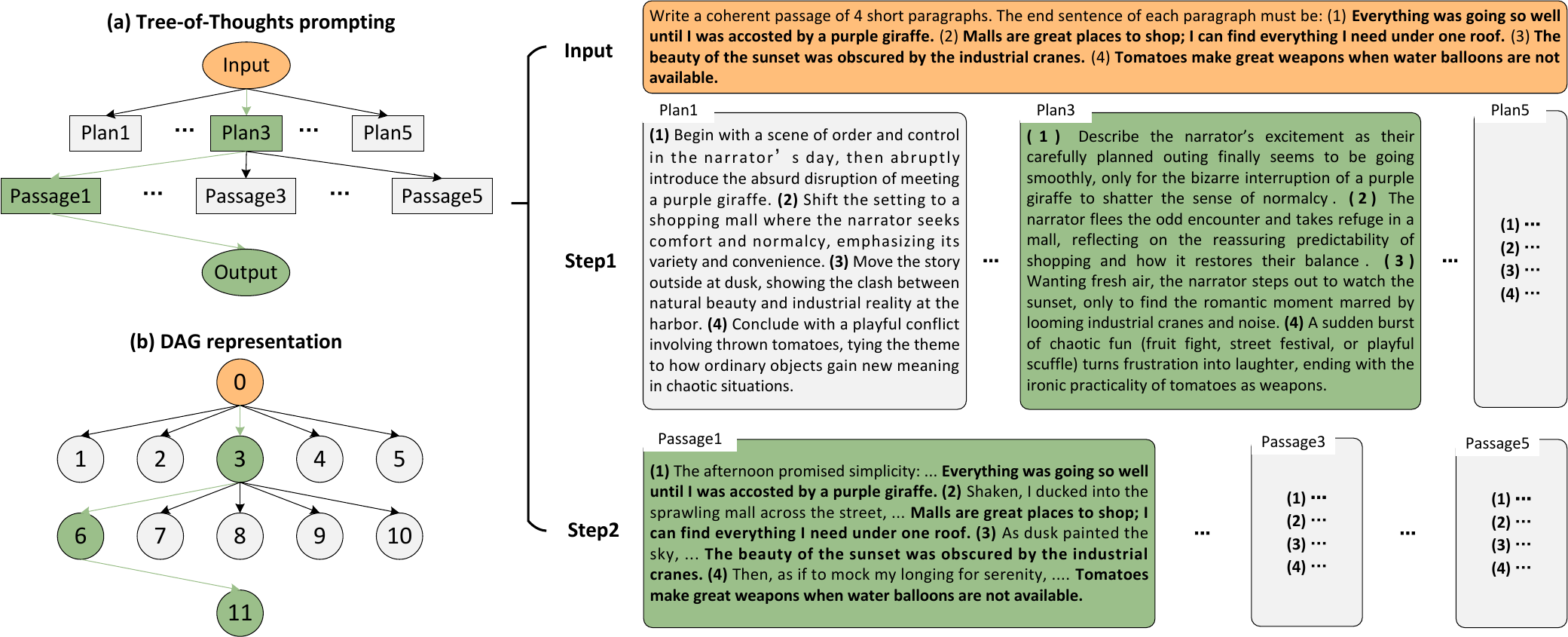}
\centering
\vspace{-1.5mm}
\caption{An example of ToT prompting in a creative writing AIGC task. Given the input, at Step 1, the GenAI model generates five different plans and selects the best one. At Step 2, the GenAI model generates five passages based on the selected plan and outputs the final AIGC result.}
\label{fig:creative writing}
\vspace{-2.5mm}
\end{figure*}

\vspace{-3mm}
\section{System Model and Problem Formulation} \label{sec:systemmodel}
\vspace{-.15mm}

In this section, we first present an overview of edge-enabled AIGC service provisioning. We then introduce a DAG model to characterize the reasoning process of ToT prompting, followed by the corresponding communication and computation models. Finally, we formulate the DAG-based thought assignment problem.

\vspace{-3mm}
\subsection{Network Outline}\label{subsec:systemarchitecture}
\vspace{-.15mm}
As shown in Fig.~\ref{fig:MEC}, we consider a network consisting of one base station (BS) equipped with an MEC server, one job owner (JO), and $U$ service providers (SPs) indexed by the set $\mathcal{U}=\{1,\ldots,U\}$. The JO generates an AIGC task,\footnote{In this paper, we consider a single JO with one AIGC task as a typical setting\cite{yan2020offloading}. The framework can be readily extended to multiple JOs with multiple AIGC tasks by assuming that AIGC tasks arrive at the BS according to a Poisson process. In addition, we use creative writing AIGC tasks as a representative example, while the same methodology can also be applied to other content types, such as images and audio (see Remark~\ref{rem:polydomain}).} but does not host a GenAI model locally and therefore requests AIGC service from the BS. The BS is equipped with a well-trained GenAI model (e.g., GPT-5.4) to deliver AIGC services using ToT prompting and acts as a centralized controller with global knowledge of the network. SPs are edge devices, such as laptops and desktop computers, each equipped with a lightweight GenAI model (e.g., Qwen3.5-Flash). These SPs can cooperatively assist the BS in delivering AIGC services.

\subsection{Tree-of-Thoughts Prompting}\label{subsec:ToT}
\vspace{-.15mm}

ToT~\cite{yao2023tree} extends the popular \emph{Chain-of-Thought}~\cite{wei2022chain} prompting by enabling exploration of coherent text units (i.e., thoughts) that serve as intermediate steps in delivering AIGC services. ToT allows GenAI models to engage in deliberate decision-making by considering multiple reasoning paths. Specifically, ToT involves four key steps: (1) decomposing the reasoning process into several steps $ToT_{\mathsf{step}}$, (2) generating potential thoughts at each step $ToT_{\mathsf{thought}}$, (3) evaluating these thoughts, and (4) selecting an appropriate search algorithm.

\vspace{-1.5mm}
\subsubsection{Thought Decomposition and DAG Modeling}
\vspace{-.15mm}

Based on the nature of the problem, ToT decomposes the reasoning process into intermediate thought steps. Depending on the task, a thought may consist of a few words, a line of an equation, or an entire paragraph of a writing plan. As an example of a creative writing AIGC task, consider providing four random sentences and requiring the output to be a coherent passage with four paragraphs, each concluding with one of the input sentences. As shown in Fig.~\ref{fig:creative writing}(a), we construct a ToT with \emph{two} steps, each containing \emph{five} thoughts. At step one, the GenAI model generates five thoughts corresponding to five writing plans and evaluates them to select \emph{one} candidate. Consequently, in the next step, ToT generates five thoughts corresponding to passages based on the best writing plan and then evaluates them to select one candidate as the final AIGC result.\footnote{Note that the number of steps $ToT_{\mathsf{step}}$, the number of thoughts generated at each step $ToT_{\mathsf{thought}}$, and the number of candidates retained per step, are hyperparameters that are typically predetermined before the ToT prompting begins.} 

To capture the intrinsic multi-step, dependency-aware, and branching characteristics of ToT prompting, we model the ToT reasoning process as a directed acyclic graph (DAG) $\mathcal{G}=(\mathcal{I},\mathcal{E})$. This representation is well aligned with the structure of ToT, where each thought depends on preceding thoughts and may lead to multiple subsequent reasoning branches. As shown in Fig.~\ref{fig:creative writing}(b), each vertex in $\mathcal{I}$ represents a thought $i$, and each directed edge $(i,j)\in \mathcal{E}$ indicates that thought $i$ must be generated before thought $j$. For clarity of exposition, we index the input and output thoughts as 0 and $|\mathcal{I}|+1$, respectively. By enforcing local generation of both the input and output thoughts, we ensure that the ToT prompting process always starts and ends at the BS.

\vspace{-1.5mm}
\subsubsection{Thought Generation and Assignment}
\vspace{-.15mm}
At each step, each thought is generated by independently issuing the same prompt to the GenAI model. Since each model call is separate and stochastic, the results are mutually independent, increasing the likelihood of exploring diverse reasoning paths. Furthermore, since different SPs are equipped with different GenAI models, these independent model calls can be executed in parallel across SPs, enabling faster thought generation and enhancing diversity through heterogeneous model capabilities.

% We define a set of binary indicators $\bm{\xi}=\{\xi_{i,0},\xi_{i,u}\mid i \in \mathcal{I},u\in \mathcal{U}\}$, where $\xi_{i,0}=1$ indicates that thought $i$ is generated locally at the BS, and $\xi_{i,0}=0$ otherwise. Similarly, $\xi_{i,u}=1$ indicates that thought $i$ is generated at UAV $u$, and $\xi_{i,u}=0$ otherwise. Since both the input and output thoughts must be generated locally, we have $\xi_{0,0} = \xi_{|\mathcal{I}|+1,0} = 1$.
Define a binary variable $x_{i,m}$, where $m \in \mathcal{U} \cup \{0\}$, to represent the \emph{thought assignment decision}, i.e., the server assigned to generate thought $i$ at each step of the ToT prompting process (e.g., a writing plan at Step 1 or a passage at Step 2, as illustrated in Fig.~\ref{fig:creative writing}(a)). Specifically, $x_{i,m}=1$ if thought $i$ is assigned to edge server $m$ for generation, where $m=0$ denotes the BS and $m=u$ represents SP $u$; otherwise, $x_{i,m}=0$. Since both the input and output thoughts must be generated locally, we have $x_{0,0} = x_{|\mathcal{I}|+1,0} = 1$. Finally, upon completion of the ToT prompting process, the BS transmits the final AIGC result back to the JO.

% For all thoughts, let $X_{0:|\mathcal{I}|+1}=[x_{0,m}, x_{1,m}, \cdots, x_{|\mathcal{I}|+1,m}]$ denote a thought assignment plan.

\vspace{-1.5mm}
\subsubsection{Thought Evaluation}
\vspace{-.15mm}
Each thought is evaluated independently through a separate call to the GenAI model using a dedicated evaluation prompt. The model is asked to assess the given thought and output a score that can be mapped to a numerical value, such as a 1–10 scale. For creative writing AIGC tasks, this approach is particularly natural, as both writing plans and generated passages are complete language segments. In such cases, the evaluation prompt can be designed to have the model rate the thought based on coherence and diversity, producing a numerical score that guides which thoughts are retained for further exploration in the ToT prompting process.

\vspace{-1.5mm}
\subsubsection{Search Algorithm}
\vspace{-.15mm}
At each step, we apply a breadth-first search strategy, retaining the most promising thought based on its evaluation score and using it as the starting point for the next expansion. This approach allows the search to explore multiple candidate paths simultaneously, preserving diversity and preventing premature convergence to a single reasoning trajectory. It should be noted that the search algorithm itself is not the optimization objective of this work; alternative strategies such as depth-first search~\cite{yao2023tree} could also be applied, and refining search algorithms is left for future work.

\vspace{-3mm}
\subsection{Communication Models}\label{subsec:communicationmodels}
\vspace{-.15mm}

Wireless transmission is required whenever two dependent thoughts are assigned to different edge servers. Specifically, if a predecessor thought is generated at the BS, i.e., $x_{i,0}=1$, and its successor is assigned to SP $u$, i.e.,  $x_{j,u}=1$, the intermediate result must be transmitted over a BS-to-SP link. Likewise, if two dependent thoughts are assigned to different SPs, i.e., $x_{i,u}=1$ and $x_{j,v}=1$ with $u \neq v$, the corresponding data must be delivered through an SP-to-SP link.

Let $a$ and $b$ denote two communicating nodes, where $(a,b)$ can represent either a BS-to-SP link or an SP-to-SP link. The achievable transmission rate (in bits per second) from node $a$ to node $b$ at time slot $t$ is given by
\vspace{-.5mm}
\begin{equation}
R_{a,b,t}=B\text{log}_2\Bigg(1+\frac{p_ah_{a,b,t}}{BN_0}\Bigg),
\end{equation}
where $B$ is the pre-allocated bandwidth (in Hz) under orthogonal frequency division multiple access among the BS and $U$ SPs, $p_a$ is the transmit power (in W) of node $a$, $h_{a,b,t}$ is the channel power gain from node $a$ to node $b$ at time slot $t$, and $N_0$ is the noise power spectrum density (in W/Hz).

The channel power gain $h_{a,b,t}$ captures both large-scale path loss and small-scale fading, and is modeled as
\vspace{-.5mm}
\begin{equation}
h_{a,b,t}=\frac{|g_{a,b,t}|^2}{10^{\text{PL}_{a,b,t}/{10}}},
\end{equation}
where $g_{a,b,t} \sim \mathcal{CN}(0,1)$ denotes the small-scale fading coefficient, which is assumed to vary independently across time slots, and $\text{PL}_{a,b,t}=127+30\text{log}_{10}(d_{a,b,t})$~\cite{8314696} is the large-scale path loss (in dB) between nodes $a$ and $b$ at time slot $t$, with $d_{a,b,t}$ denoting the corresponding link distance (in meter).

\vspace{-3mm}
\subsection{Computing Models}\label{subsec:computingmodels}
\vspace{-.15mm}

Before introducing our computing models, we first highlight the major challenges in evaluating the performance of AIGC services. Unlike most MEC studies~\cite{wang2021dependent,yan2020offloading,nguyen2023dependency,zhou2023dag}, where computational resources are clearly defined as CPU cycles per second, the notion of computational resources in GenAI models remains ambiguous. Moreover, unlike traditional performance indicators such as task execution delay and energy consumption, the evaluation metrics for AIGC services remain unclear. It is essential to identify which controllable parameters truly influence the generation time and quality of AIGC outputs. Without a well-defined expression linking GenAI computational resources to both generation delay and output quality, it is impossible to quantify AIGC service provisioning.

% running Ubuntu 20.04.6, Python 3.10, PyTorch 2.5.0, and CUDA 12.0,

To address this, we focus on a specific type of AIGC--creative writing--and define the computational resources of GenAI models in terms of the number of output tokens.\footnote{Considering that for generation time, a longer output length requires more decoding steps, causing the generation time to increase approximately linearly with the number of output tokens. On the other hand, for generation quality, a longer output length typically enables the language model to produce more semantically rich, coherent, and well-structured content in creative writing AIGC tasks. Therefore, we treat the number of output tokens as the computational resource in GenAI models in this paper.} We then sample 40 random sentences from \url{https://randomwordgenerator.com/sentence.php} and deploy Qwen 2.5-7B-Instruct~\cite{qwen2.5}, an open-source language model with outstanding performance, on a platform equipped with an NVIDIA 4090 GPU. Consequently, we conduct extensive experiments (detailed in Sec.~\ref{subsubsec:generatequality} and Sec.~\ref{subsubsec:generatedelay}) and apply parameter fitting methods to establish one of the first mathematical relationships among output token count, writing generation delay, and the quality of the generated text.

\begin{figure}[t!]
\includegraphics[width=.48\textwidth]{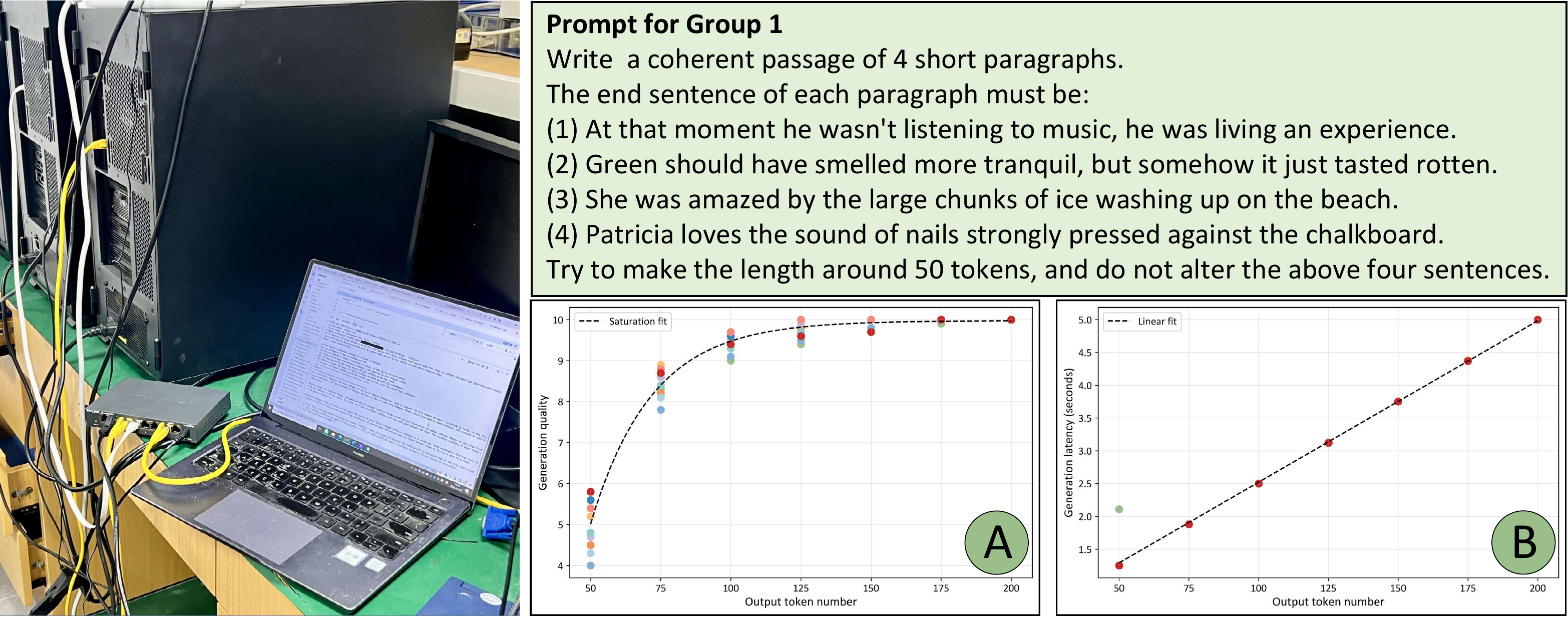}
\centering
\vspace{-1.5mm}
\caption{{Hardware setup for providing a creative-writing AIGC service with different output token counts. The server is equipped with an NVIDIA 4090 GPU, on which the open-source large language model Qwen2.5-7B-Instruct is deployed.}}
\label{fig:fitting function}
%\vspace{-1mm}
\end{figure}

\vspace{-1.5mm}
\subsubsection{Generation Quality}\label{subsubsec:generatequality}
\vspace{-.15mm}
Given that passage quality is inherently subjective and no reference sentences are available, direct evaluation is highly challenging. One possible approach is to rely on human judgments, such as hiring professional writers to assess the quality of the generated passages, but this method is highly costly. To this end, considering that GPT-5.4 has been trained on vast amounts of text and has learned general patterns of coherence, fluency, and relevance~\cite{van2023chatgpt}, we evaluate the generated passages using a GPT-5.4 zero-shot prompt that provides a scalar score from 1 to 10. In this setting, GPT-5.4 generates the score without examples, where a higher score indicates better sentence quality.

We divide 40 random sentences into 10 groups as 10 creative writing AIGC tasks. In each task, the input consists of 4 sentences, and the output is required to be a coherent passage with 4 paragraphs, each ending with one of the input sentences. For each creative writing AIGC task, we input the corresponding 4 sentences into Qwen 2.5-7B-Instruct, and evaluate the generated passages using a zero-shot prompt that assigns a scalar score from 1 to 10. Since we define the computational resources of GenAI models in terms of output token count, we vary the number of output tokens in Qwen 2.5-7B-Instruct, selecting values from the set $\{50, 75, 100, 125, 150, 175, 200\}$. As shown in Part A of Fig.~\ref{fig:fitting function}, we present the scores of the 10 creative writing AIGC tasks under different output token counts and fit them using an exponential saturation function.\footnote{Since ToT prompting consists of multiple thoughts generated and evaluated independently, in this experiment we consider only a single-step passage generation. We treat this process as one thought, from which the corresponding mathematical expression is derived.}

As a result, we propose a general mathematical relationship that links the score of generating thought $i$ on server $m$ (in this experiment, $m$ can represent any server equipped with Qwen 2.5-7B-Instruct), denoted by $Score_{i,m}$, to the output token count $C_m$, expressed as follows:
\vspace{-.5mm}
\begin{equation}\label{eq:AIGCquality}
Score_{i,m}=Score_{max}-\sigma_{m}e^{-\rho_{m} C_m},
\end{equation}
where, in this experiment, $Score_{max}=10$ denotes the maximum score, $\sigma_{m} = 49.13$ represents the initial score deficit (a larger value indicates that the model starts from a lower baseline and therefore requires more tokens to approach the performance ceiling), and $\rho_{m} = 0.046$ denotes the rate of quality improvement as the output token count increases (a larger value indicates faster convergence).

% We then construct a ToT with $c=2$ steps, where Qwen 2.5-7B-Instruct first generates $k=5$ thoughts at Step 1 corresponding to five writing plans and evaluates them to select $b=1$ candidate. Subsequently, at Step 2, based on the best plan, Qwen 2.5-7B-Instruct generates $k=5$ thoughts corresponding to five passages and evaluates them to select $b=1$ candidate as the final AIGC result.
\vspace{-1.5mm}
\subsubsection{Generation Delay}\label{subsubsec:generatedelay}
\vspace{-.15mm}
Using the same methodology, we report the generation times of the 10 creative writing AIGC tasks under different output token counts and fit them with a function that appears linear, as shown in Part B of Fig.~\ref{fig:fitting function}. Similarly, we propose a general mathematical relationship that links the generation delay of thought $i$ on server $m$, denoted by $T^{\mathsf{gen}}_{i,m}$, to the output token count $C_m$, expressed as follows:
\vspace{-.5mm}
\begin{equation}\label{eq:AIGCdelay}
T^{\mathsf{gen}}_{i,m}=\eta_{m}C_m+\psi_{m},
\end{equation}
where, in this experiment, $\eta_{m} = 0.025$ represents the generation delay per token (a larger value indicates higher latency per token), and $\psi_{m} = 0.062$ denotes the initial generation overhead (a larger value indicates a longer base delay).

\vspace{-.5mm}
\begin{remark}\label{rem:polydomain} While our experiments focus on creative writing as a representative AIGC task, the proposed methodology is not limited to this specific scenario. In creative writing, the number of output tokens is closely related to both the expressiveness and coherence of the generated text, making it a natural proxy for computational resources. However, for other types of AIGC tasks, alternative controllable inference parameters (e.g., reasoning depth or search beam width) can be used to characterize computational resource consumption. By fitting the corresponding mathematical relationships and integrating them into the optimization model, the proposed ToT-empowered AIGC service framework can be readily extended to a broad range of reasoning and generation tasks beyond creative writing.
\end{remark}
\vspace{-.5mm}

Next, to characterize the execution time of each thought (including transmission and generation) in the local BS and SPs, we first define the notions of \emph{ready time} and \emph{finish time}.

\vspace{-.5mm}
\begin{definition}[Finish Time]\label{def:finishtime}
The finish time of thought $i$ refers to the moment when it has been generated. We denote $T^{\mathsf{fin}}_{i,m}$ as the finish time when thought $i$ is generated at server $m$, where $m \in \mathcal{U} \cup \{0\}$.
\end{definition}
\vspace{-.5mm}

\begin{figure}[t!]
\includegraphics[width=.4\textwidth]{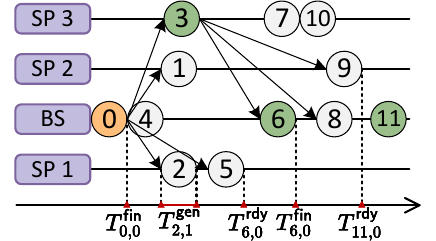}
\centering
\vspace{-1.5mm}
\caption{{An illustration of the thought assignment and generation timeline based on the DAG structure in Fig.~\ref{fig:creative writing}(b), with $U=3$ SPs. Each row represents a server (BS, SP~1, SP~2, SP~3), and each block denotes the generation of a specific thought.}}
\label{fig:timeline}
%\vspace{-1mm}
\end{figure}

\vspace{-.5mm}
\begin{definition}[Ready Time]\label{def:readytime}
The ready time of a thought is the earliest moment when it has received all the necessary input data (e.g., prompting information) to begin generation. For example, in Fig.~\ref{fig:creative writing}(b), the ready time of thought $6$ is the time when the input data stream from thought $3$ has arrived. We denote the ready time of thought $i$ generated at server $m$ as $T^{\mathsf{rdy}}_{i,m}$. To further illustrate these concepts, Fig.~\ref{fig:timeline} presents the thought assignment and generation timeline corresponding to the DAG structure in Fig.~\ref{fig:creative writing}(b), from which the ready time and finish time of each thought across different servers can be clearly observed.
\end{definition}
\vspace{-.5mm}

\begin{itemize}[leftmargin=4mm]
\item \emph{\textbf{Local Generation:}} We assume that the BS is equipped with a well-trained GenAI model (e.g., GPT-5.4) with an output token count of $C_0$. If thought $i$ is generated locally at the BS, its local generation time, based on~(\ref{eq:AIGCdelay}), is given by
\vspace{-.5mm}
\begin{equation}\label{eq:localtime}
T^{\mathsf{gen}}_{i,0}=\eta_{0}C_0+\psi_{0},
\end{equation}
where $\eta_{0}$ and $\psi_{0}$ are the generation delay parameters associated with the GenAI model deployed at the BS (to be provided in Sec.~\ref{sec:simulation}), and the corresponding generation quality, based on~(\ref{eq:AIGCquality}), is calculated as
\vspace{-.5mm}
\begin{equation}\label{eq:localquality}
Score_{i,0}=10-\sigma_{0}e^{-\rho_{0} C_0},
\end{equation}
where $\sigma_{0}$ and $\rho_{0}$ are the generation quality parameters associated with the GenAI model deployed at the BS (to be specified in Sec.~\ref{sec:simulation}). Then, the ready time $T^{\mathsf{rdy}}_{i,0}$ of thought $i$ generated locally at the BS is given by
\vspace{-.5mm}
\begin{align}
\hspace{-1.5mm} T^{\mathsf{rdy}}_{i,0}&=\max_{j \in \mathcal{L}_i-1}\Bigg\{ 
T_{j}^{\mathsf{actfin}} x_{j^*,0} \nonumber\\
&+\sum_{u=1}^{U}
\left(
T_{j}^{\mathsf{actfin}} + \frac{e_{j^*,i}}{R_{u,0,t_{i}}}
\right)x_{j^*,u}\Bigg\}.
\end{align}
% \begin{align}
% \hspace{-1.5mm} T^{\mathsf{rdy}}_{i,0}= \max_{j \in \mathcal{P}_i}\Bigg\{
% T_{j,0}^{\mathsf{fin}} x_{j,0}+
% \sum_{u=1}^{U}
% \left(
% T_{j,u}^{\mathsf{fin}} + \frac{e_{j,i}}{R_{u,0,t_j}}
% \right)\!x_{j,u}\! \Bigg\}.
% \end{align}
Here, $\mathcal{L}_i$ denotes the step at which thought $i$ is located (e.g., in Fig.~\ref{fig:creative writing}, we have $\mathcal{L}_3=1$ and $\mathcal{L}_6=2$), $j^*=\text{argmax}_{j\in \mathcal{L}_i-1}Score_{j,m}x_{j,m}$ denotes the thought that obtains the highest score at step $\mathcal{L}_i-1$, $T_{j}^{\mathsf{actfin}}$ denotes the actual completion time of thought $j$ when generated on the designated server, and the $\text{max}$ operator guarantees that thought evaluation can only be performed after all thoughts have been generated. Specifically, if $x_{j^*,0}=1$ for thought $j^*$, the time until its output data becomes available at the BS is equal to the latest local finish time $\max_{j \in \mathcal{L}_i-1}T_{j}^{\mathsf{actfin}}$. Otherwise, if $x_{j^*,u} = 1$, the time until its output data becomes available at the BS for generating thought $i$ equals the latest completion time $\max_{j \in \mathcal{L}_i-1}T_{j}^{\mathsf{actfin}}$ plus the data transmission time $\frac{e_{j^*,i}}{R_{u,0,t_i}}$, where $e_{j^*,i}$ denotes the input data size associated with edge $(j^*,i)$. Additionally, $t_i=\max_{j \in \mathcal{L}_i-1}T_{j}^{\mathsf{actfin}}$ denotes the time slot at which thought $i$ begins receiving the corresponding input data. When all required data is available at the ready time $T^{\mathsf{rdy}}_{i,0}$, the BS generates thought $i$ with the generation time $T^{\mathsf{gen}}_{i,0}$ from~(\ref{eq:localtime}), so the finish time of thought $i$ for local generation is given by
\vspace{-.5mm}
\begin{equation}
T_{i,0}^{\mathsf{fin}}=\max \big\{T^{\mathsf{rdy}}_{i,0}, A_0\big\}+T^{\mathsf{gen}}_{i,0},
\end{equation}
where $A_0$ denotes the time when the BS finishes generating its most recently assigned thought and becomes available to generate thought $i$.

\item \emph{\textbf{MEC Generation:}} We assume that each SP $u$ is equipped with a lightweight GenAI model (e.g., Qwen3.5-Flash) with an output token count of $C_{u,t}$ at time slot $t$\footnote{Unlike the BS, which is assumed to possess fixed and ample computing resources, each SP has limited and time-varying computational availability due to its constrained hardware capacity and the need to concurrently support its own local workloads.} ($C_{u,t} < C_0, \forall u \in \mathcal{U}$). If thought $i$ is generated at SP $u$ during time slot $t$, its generation time, based on~(\ref{eq:AIGCdelay}), is given by
\vspace{-.5mm}
\begin{equation}\label{eq:mectime}
T^{\mathsf{gen}}_{i,u,t}=\eta_{u}C_{u,t}+\psi_{u},
\end{equation}
where $\eta_{u}$ and $\psi_{u}$ are the generation dalay parameters associated with the GenAI model deployed at SP $u$ (to be provided in Sec.~\ref{sec:simulation}), and the corresponding generation quality, based on~(\ref{eq:AIGCquality}), is calculated as
\vspace{-.5mm}
\begin{equation}\label{eq:mecquality}
Score_{i,u}=10-\sigma_{u}e^{-\rho_{u} C_u},
\end{equation}
where $\sigma_{u}$ and $\rho_{u}$ are the generation quality parameters associated with the GenAI model deployed at SP $u$ (to be specified in Sec.~\ref{sec:simulation}). Similarly, the ready time of thought $i$ generated at SP $u$ can be calculated as
\vspace{-.5mm}
\begin{align}
T^{\mathsf{rdy}}_{i,u}&= \max_{j \in \mathcal{L}_i-1}\Bigg\{T_{j}^{\mathsf{actfin}}x_{j^*,u}+\Bigg(T_{j}^{\mathsf{actfin}} +\frac{e_{j^*,i}}{R_{0,u,t_i}}\Bigg)x_{j^*,0}  \nonumber\\
&+\sum_{v=1,v\neq u}^U \Bigg(T_{j}^{\mathsf{actfin}} 
+\frac{e_{j^*,i}}{R_{v,u,t_i}}\Bigg)x_{j,v}\Bigg\},
\end{align}
where $t_i=\max_{j \in \mathcal{L}_i-1}T_{j}^{\mathsf{actfin}}$ denotes the time slot at which thought $i$ begins receiving the corresponding input data. Finally, the finish time of thought $i$ generated at SP $u$ is given by
\vspace{-.5mm}
\begin{equation}
T_{i,u}^{\mathsf{fin}}=\max \{T^{\mathsf{rdy}}_{i,u}, A_u\}+T^{\mathsf{gen}}_{i,u,\max \{T^{\mathsf{rdy}}_{i,u}, A_u\}}.
\end{equation}
\end{itemize}

\vspace{-3mm}
\subsection{Problem Formulation}\label{subsec:probformulate}
\vspace{-.15mm}
From the above discussion, considering the interdependencies among different thoughts and the requirement that both the input and output thoughts must be generated locally at the BS, the total finish time to complete the ToT prompting is equal to the local finish time of the output thought, i.e., $T_{tot} = T_{|\mathcal{I}|+1,0}^{\mathsf{fin}}$. Moreover, the total generation quality of the ToT prompting can be calculated as\footnote{In this work, we recognize that each thought’s quality influences subsequent ones and ultimately the final result. Therefore, we evaluate the quality of all thoughts throughout the ToT prompting process.}
\vspace{-.5mm}
\begin{equation}
Score_{tot}=\sum_{i=0}^{{|\mathcal{I}|+1}}\Bigg(Score_{i,0}x_{i,0}+\sum_{u=1}^UScore_{i,u}x_{i,u}\Bigg).
\end{equation}

To deliver low-latency AIGC services,\footnote{In the era of large language models, where a single inference already demands considerable time, the multi-step structure of ToT compounds this delay significantly, leading to unacceptable waiting times for end users.} we minimize the total generation delay subject to a user-adjustable quality constraint:
\vspace{-1.5mm}
\begin{align}
     &(\textbf{P}) \quad\quad\quad \min_{\mathcal{X}} \quad T_{tot},\hspace{-40mm}\label{eq:problem1} \nonumber \\
     \text{s.t.} \quad
     & \mathcal{C}1: x_{i,m} \in \{0,1\}, \ \forall i\in \mathcal{I}, m \in \mathcal{U} \cup\{0\},  \nonumber \\
     &\mathcal{C}2: \ Score_{tot} \geq Score_{min},  \nonumber 
\end{align}
where $\mathcal{X}=[x_{i,m}]_{|\mathcal{I}| \times (|\mathcal{U}|+1)}$ is a binary matrix representing the complete thought assignment plan. Constraint $\mathcal{C}1$ enforces the binary nature of the assignment variables, and constraint $\mathcal{C}2$ ensures that the overall generation quality meets a minimum threshold $Score_{min}$ specified by the user.

\vspace{-.5mm}
\begin{remark}\label{rem:NPhard} Due to the interdependencies among thoughts in the ToT prompting process--where each thought’s generation depends on the outputs of its predecessor(s)--the optimization problem exhibits strong temporal and structural coupling. Moreover, the presence of the binary variable $x_{i,m}$ makes the problem inherently discrete. Consequently, problem $(\textbf{P})$ is non-convex and challenging to solve.
\end{remark}
\vspace{-.5mm}

\vspace{-3mm}
\section{Overview of Diffusion Models} \label{sec:diffusionmodel}
\vspace{-.15mm}
Before introducing our diffusion-based soft actor-critic (DSAC) algorithm, we first present the challenges faced by conventional DRL algorithms, which serve as the main motivation for integrating diffusion models with DRL. We then present the adaptation of the diffusion model specifically designed to generate optimal thought assignment decisions.

\vspace{-3mm}
\subsection{Motivation of Adopting Diffusion Model}
\vspace{-.15mm}
Since the thought assignment decisions in this paper are binary, we primarily focus on value-based DRL algorithms (e.g., DQN) and their inherent limitations. Specifically, value-based DRL derives a deterministic policy by updating the deep Q-network based on one-step rewards. However, this update directly affects only the value of the state–action pair that produced the reward, while the values of other pairs are influenced only indirectly through the updated Q-network~\cite{mnih2016asynchronous}. Considering the rapidly varying channel conditions, the number of state–action pairs becomes substantial. Consequently, the learned Q-values for different thought assignment decisions cannot be efficiently generalized, resulting in slow convergence, inefficient exploration, and suboptimal performance.

The denoising diffusion probabilistic model (DDPM)~\cite{ho2020denoising}, originally developed for image generation, inspires our approach to address the shortcomings mentioned above. Specifically, in a standard DDPM implementation, the training involves two key stages: \emph{1) the forward process}, which gradually adds noise sampled from a standard Gaussian distribution to the input image over multiple steps until it becomes indistinguishable from isotropic Gaussian noise; and \emph{2) the reverse process}, in which a neural network learns to progressively remove the noise step by step to reconstruct the original image. Building on the iterative refinement mechanism and strong generative capabilities of DDPM, it can gradually refine noisy latent representations into high-quality action candidates, rather than relying on single-step updates. Additionally, our adoption of diffusion models is further motivated by their strong compatibility with DRL frameworks~\cite{liu2024dnn,11477959}. Specifically, in a conventional diffusion model, a user can input a text prompt (e.g., `an apple on the table') to guide the model in generating a corresponding image. In our scenario, we treat the system state (e.g., channel conditions) as the 'input text prompt' and define the optimal thought assignment decisions as the `target image' to be generated.

% in the reverse process of diffusion models (detailed in Sec.~\ref{subsubsec:reverse_process}), a user-provided text prompt is used to guide the denoising process, generating images that align with the user's preferences (detailed in Sec.~\ref{subsec:DDPM}). 

\vspace{-3mm}
\subsection{Preliminaries of Diffusion Models}  \label{subsec:DDPM}
\begin{figure}[t]
\includegraphics[width=.48\textwidth]{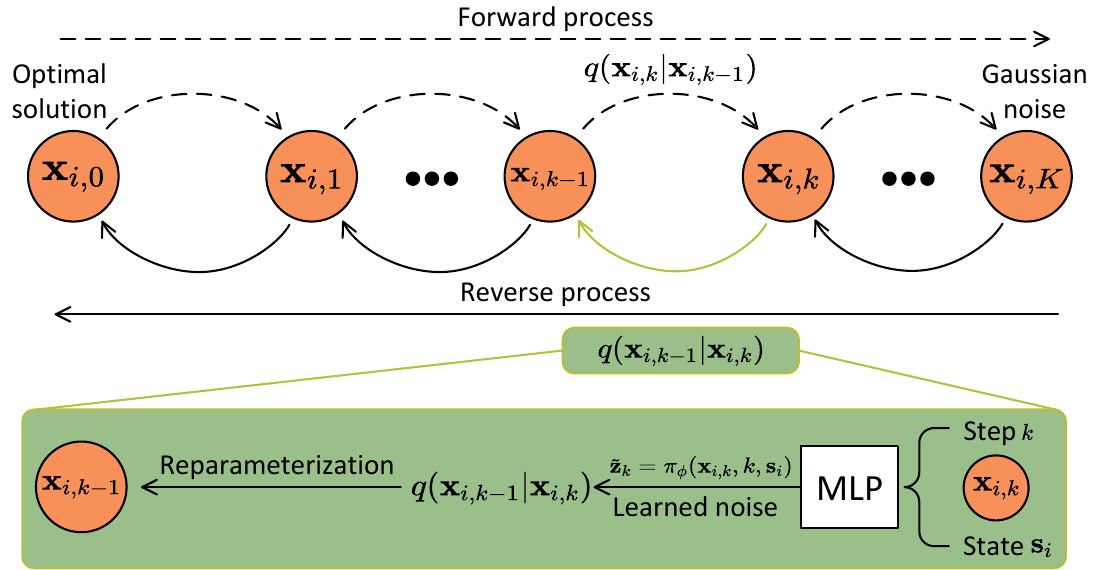}
\centering
\vspace{-1.5mm}
\caption{An illustration of the diffusion model tailored to generate the optimal assignment decision for thought $i$.}
\label{fig:DDPM}
\end{figure}

\vspace{-.15mm}
For thought $i$, we represent its optimal assignment decision $\mathbf{x}_{i,0}$\footnote{We make a slight adjustment to the thought assignment decision in our simulations, where $x_{i,m}\in \{0,1\}$ is extended to $\mathbf{x}_{i}\in \{0,1,\ldots,U\}$. Here, $\mathbf{x}_{i}=u$ indicates that $x_{i,u}=1$ within our proposed system model.} as a set of discrete probabilities over the choice of being generated at the BS or one of the SPs, i.e., $\mathbf{x}_{i,0} \sim \mathbb{R}^{|\mathcal{U}|+1}$. According to the diffusion model, the optimal assignment decision $\mathbf{x}_{i,0}$ can be progressively perturbed with noise until it becomes Gaussian, a process known as \emph{the forward process}. Then, in \emph{the reverse process}, the denoiser, denoted by $\pi_{\phi}$ and parameterized by $\phi$, starts from Gaussian noise and progressively reconstructs $\mathbf{x}_{i,0}$. Fig.~\ref{fig:DDPM} illustrates our diffusion model framework for generating the optimal assignment decision for thought $i$. In the following, we present the forward and reverse processes, respectively.

\vspace{-1.5mm}
\subsubsection{The Forward Process}  \label{subsubsec:forwardprocess}  
\vspace{-.15mm}
Given the optimal assignment decision $\mathbf{x}_{i,0}$, the forward process adds a sequence of Gaussian noise at each step $k$ to obtain $\mathbf{x}_{i,1}, \mathbf{x}_{i,2}, \dots, \mathbf{x}_{i,K}$, where $\mathbf{x}_{i,k}$ is the discrete vector of the distribution at step $k$ and has the same dimensionality as $\mathbf{x}_{i,0}$. The transition from $\mathbf{x}_{i,k-1}$ to $\mathbf{x}_{i,k}$ is modeled as a normal distribution with a mean of $\sqrt{1-\beta_k}\mathbf{x}_{i,k-1}$ and a variance of $\beta_k\mathbf{I}$, as given in~\cite{ho2020denoising}.
\vspace{-.5mm}
\begin{equation}\label{eq:forwarddistribution}
    q(\mathbf{x}_{i,k}|\mathbf{x}_{i,k-1})= \mathcal{N}(\mathbf{x}_{i,k};\sqrt{1-\beta_k}\mathbf{x}_{i,k-1},\beta_k\mathbf{I}),
\end{equation}
where $k = 1, 2, \dots, K$, $\beta_k = 1 - e^{-\frac{\beta_{min}}{K} - \frac{2k - 1}{2K^2}(\beta_{max} - \beta_{min})}$ is the diffusion rate, with $\beta_{min}$ and $\beta_{max}$ denoting the minimum and maximum rates, respectively, and $\mathbf{I}$ representing the identity matrix.

We then express the connection between $\mathbf{x}_{i,k-1}$ and $\mathbf{x}_{i,k}$ by rewriting \eqref{eq:forwarddistribution} using the reparameterization technique, as follows~\cite{ho2020denoising}:
\vspace{-.5mm}
\begin{equation}\label{eq:forwardupdate}
    \mathbf{x}_{i,k}= \sqrt{1-\beta_k}\mathbf{x}_{i,k-1}+\sqrt{\beta_k}\mathbf{z}_{k-1},
\end{equation}
where $\mathbf{z}_{k-1}$ is standard Gaussian noise sampled from $\mathcal{N}(0,\mathbf{I})$. Consequently, the mathematical relationship between $\mathbf{x}_{i,0}$ and $\mathbf{x}_{i,k}$ at any step $k$ can be derived as:
\vspace{-.5mm}
\begin{align}\label{eq:forwardrelationship}
    \mathbf{x}_{i,k}=\sqrt{\bar{\alpha}_k}\mathbf{x}_{i,0}+\sqrt{1-\bar{\alpha}_k}{\mathbf{z}}_k,
\end{align}
where $\alpha_{k}=1-\beta_{k}$, $\bar{\alpha}_k=\alpha_{1}*\alpha_{2}*\ldots*\alpha_{k}$, and ${\mathbf{z}}_k \sim \mathcal{N}(0,\mathbf{I})$ denotes standard Gaussian noise.

\vspace{-1mm}
\begin{remark} Since our goal is to obtain the optimal assignment decision $\mathbf{x}_{i,0}$ for thought $i$ using the diffusion model, the forward process--which requires $\mathbf{x}_{i,0}$ in advance to serve as the `original image' to be progressively perturbed with noise--contradicts our objective. Therefore, the forward process is omitted in this work, as indicated by the dotted lines in Fig.~\ref{fig:DDPM}. That is, the forward process here primarily defines the mathematical relationship between $\mathbf{x}_{i,0}$ and $\mathbf{x}_{i,k}$, providing the necessary foundation for the subsequent reverse process.
\end{remark}
\vspace{-.5mm}

\vspace{-1.5mm}
\subsubsection{The Reverse Process}  \label{subsubsec:reverseprocess} 
\vspace{-.15mm}
Since $\mathbf{x}_{i,k}$ becomes indistinguishable from isotropic Gaussian noise as $K$ increases, in the reverse process we initialize with $\mathbf{x}_{i,K} \sim \mathcal{N}(0, \mathbf{I})$ and iteratively remove the noise to reconstruct the optimal thought assignment decision $\mathbf{x}_{i,0}$. Therefore, the core of the reverse process is to determine the transition from $\mathbf{x}_{i,k}$ to $\mathbf{x}_{i,k-1}$, which cannot be calculated directly. However, it follows a Gaussian distribution as given in~\cite{ho2020denoising}:
\vspace{-.5mm}
\begin{align}\label{eq:reversedistribution}
q(\mathbf{x}_{i,k-1}|\mathbf{x}_{i,k})= \mathcal{N}(\mathbf{x}_{i,k-1};\tilde{\mu}_{i,k},\tilde{\beta}_k\mathbf{I}),
\end{align} 
where $\tilde{\beta}_k = \frac{1 - \bar{\alpha}_{k-1}}{1 - \bar{\alpha}_k}\beta_k$ is the variance amplitude, which can be easily computed from deterministic parameters, and $\tilde{\mu}_{i,k}$ can be derived through Bayesian inference~\cite{ho2020denoising}:
\vspace{-.5mm}
\begin{align}\label{eq:originalmean}
\tilde{\mu}_{i,k}= \frac{\sqrt{\alpha_k}(1-\bar{\alpha}_{k-1})}{1-\bar{\alpha}_k}\mathbf{x}_{i,k}+\frac{\sqrt{\bar{\alpha}_{k-1}}\beta_k}{1-\bar{\alpha}_k}\mathbf{x}_{i,0}.
\end{align}

Next, by substituting the mathematical relationship between $\mathbf{x}_{i,0}$ and $\mathbf{x}_{i,k}$ obtained in~\eqref{eq:forwardrelationship} into~\eqref{eq:originalmean}, we can reformulate the mean $\tilde{\mu}_{i,k}$ as:
\vspace{-.5mm}
\begin{align}
\tilde{\mu}_{i,k}=\frac{1}{\sqrt{\alpha_k}}\Bigg(\mathbf{x}_{i,k}-\frac{1-\alpha_k}{\sqrt{1-\bar{\alpha}_k}}\tilde{\mathbf{z}}_k\Bigg),
\end{align}
where $\tilde{\mathbf{z}}_k$ is a new source of noise at each step $k$, independent of the noise $\mathbf{z}_k$ added during the forward process. To obtain $\tilde{\mathbf{z}}_k$, we then employ the denoiser $\pi_{\phi}$, which takes three inputs: $\mathbf{x}_{k}$, the step index $k$, and the system state $\mathbf{s}_i$ when assigning thought $i$ (defined later in Sec.~\ref{subsubsec:DSACMDP}), to predict the noise to be subtracted. We further express the mean $\tilde{\mu}_{i,k}$ as:
\vspace{-.5mm}
\begin{align}\label{eq:reversemean}
\tilde{\mu}_{i,k}=\frac{1}{\sqrt{\alpha_k}}\Bigg(\mathbf{x}_{i,k}-\frac{1-\alpha_k}{\sqrt{1-\bar{\alpha}_k}}\pi_{\phi}(\mathbf{x}_{i,k},k,\mathbf{s}_i)\Bigg),
\end{align}

Finally, after obtaining $\tilde{\mu}_{i,k}$ and $\tilde{\beta}_k$, based on~\eqref{eq:reversedistribution}, we derive the transition from $\mathbf{x}_{i,k}$ to $\mathbf{x}_{i,k-1}$ through reparameterization:
\vspace{-.5mm}
\begin{equation}\label{eq:reverseupdate}
    \mathbf{x}_{i,k-1}\!=\! \frac{1}{\sqrt{\alpha_k}}\Bigg(\!\mathbf{x}_{i,k}\!-\!\frac{1-\alpha_k}{\sqrt{1-\bar{\alpha}_k}}\pi_{\phi}(\mathbf{x}_{i,k},k,\mathbf{s}_i)\!\!\Bigg)\!+\!\sqrt{\tilde{\beta}_k}\bar{\mathbf{z}}_k,
\end{equation}
where $\bar{\mathbf{z}}_k$ is standard Gaussian noise sampled from $\mathcal{N}(0,\mathbf{I})$. By iteratively applying the transition in~\eqref{eq:reverseupdate} over $K$ steps (as detailed in Algorithm~\ref{algo:DSAC}), we can reconstruct the optimal assignment decision $\mathbf{x}_{i,0}$ for thought $i$.

\vspace{-1mm}
\begin{remark}\label{rem:noforward}
In standard DDPM implementations, the training objective is to minimize the mean squared error (MSE) between the noise $\mathbf{z}_k$ sampled from $\mathcal{N}(0,\mathbf{I})$ in the forward process and the noise $\tilde{\mathbf{z}}_k$ predicted by the denoiser $\pi_{\phi}$ in the reverse process at each step $k$. However, since we omit the explicit forward process, we optimize the reverse process in an exploratory manner. Specifically, the training objective of the denoiser $\pi_{\phi}$ shifts from minimizing the MSE with labeled data to maximizing the objective function of problem $(\textbf{P})$ (as detailed in Sec.~\ref{sec:DSAC}).
\end{remark}
\vspace{-.5mm}

\vspace{-3mm}
\section{Diffusion-Based Soft Actor-Critic Algorithm}\label{sec:DSAC}
\vspace{-.15mm}
In this section, we first present the motivation for integrating the diffusion model into the soft actor-critic (SAC) algorithm. We then define the Markov decision process (MDP) elements in the DSAC algorithm, followed by an overview of the DSAC architecture. Finally, we provide a comprehensive analysis of its computational complexity.

\vspace{-3mm}
\subsection{Motivation for Adopting the SAC Algorithm}
\vspace{-.15mm}

The motivation for adopting the SAC algorithm stems from its superior stability, efficiency, and exploration capability. SAC is a DRL framework built on the principle of maximum entropy, where the entropy represents the randomness of the agent’s policy~\cite{haarnoja2018soft}. Specifically, traditional reinforcement learning algorithms, such as DQN and DDPG, aim to maximize only the expected cumulative reward. However, this reward-only objective often results in insufficient exploration and unstable policy learning, particularly in high-dimensional or non-stationary environments. In contrast, the SAC algorithm maximizes both the expected cumulative reward and the policy entropy, encouraging the agent to sustain sufficient exploration throughout the training process. This entropy-regularized objective allows SAC to achieve more stable learning and effectively avoid suboptimal convergence.

Moreover, SAC can be naturally extended to discrete action spaces~\cite{li2024multi}, making it well-suited for decision-making tasks such as thought assignment, where the action space is discrete yet demands efficient and robust exploration. This makes SAC particularly effective in dynamic MEC networks.

\vspace{-3mm}
\subsection{MDP Elements in the DSAC Algorithm} \label{subsubsec:DSACMDP}
\vspace{-.15mm}
Recall that we have a set of thoughts $\mathcal{I}$, and the objective is to assign each thought to either the BS or one of the SPs to maximize the QoS of AIGC service delivery. This makes the MDP framework particularly suitable for the thought assignment problem~\cite{du2024diffusion}. An MDP is a discrete-time stochastic control process that can be represented as a 3-tuple $(\mathcal{S}, \mathcal{A}, \mathcal{R})$, where $\mathcal{S}$ denotes the state space, $\mathcal{A}$ represents the action space, and $\mathcal{R}$ is the reward, as described below.

\begin{itemize}[leftmargin=4mm]
\item \textbf{\textit{State Space:}} The state space $\mathcal{S}$ encapsulates the environmental information required for making thought assignment decisions. The state $\mathbf{s}_i \in \mathcal{S}$, corresponding to the assignment of thought $i$, is defined as follows:
\vspace{-.5mm}
\begin{align}\label{eq:DSACstate}
    \mathbf{s}_i=\{\mathbf{g}_{t_i},\mathbf{C}_{t_i},\mathbf{X},e_{j^*,i} \}, \ t_i=\max_{j \in \mathcal{L}_i-1}T_{j}^{\mathsf{actfin}}, 
\end{align}
where $\mathbf{g}_{t_i}=\big\{|g_{a,b,t_i}|^2 \big\}_{a\in\mathcal{U}\cup \{0\},b\in\mathcal{U}\cup \{0\}}$ denotes the small-scale fading power gains of all BS-to-SP and SP-to-SP links at time slot $t_i$, $\mathbf{C}_{t_i}=\big\{C_{u,t_i} \big\}_{u\in\mathcal{U}}$ represents the output token counts of all SPs at time slot $t_i$, $\mathbf{X}$ denotes the assignment decisions $\mathbf{x}_i$ for all thoughts, where the vector is padded with -1 for thoughts that have not yet been assigned, and $e_{j^*,i}$ represents the input data size required to generate thought $i$, where $j^*=\text{argmax}_{j\in \mathcal{L}_i-1}Score_{j,m}x_{j,m}$ denotes the thought that obtains the highest score at step $\mathcal{L}_i-1$.

\item \textbf{\textit{Action Space:}} The action space $\mathcal{A}$ is defined as the set of all possible thought assignment decisions, i.e., $\mathcal{A}=\{0,1,2,\ldots,U\}$. Recalling that $\mathbf{x}_{i,0}$ represents a set of discrete probabilities over the choices of being generated at the BS or one of the SPs, i.e., $\mathbf{x}_{i,0} \sim \mathbb{R}^{|\mathcal{U}|+1}$, the action $\mathbf{a}_i \in \mathcal{A}$ corresponding to the assignment of thought $i$ can be expressed as
\begin{align}\label{eq:DSACaction}
    \mathbf{a}_i=\arg \max \{\mathbf{x}_{i,0}\}.
\end{align}

\item \textbf{\textit{Reward Function:}} The reward $r_i\in \mathcal{R}$ is a scalar representing the immediate feedback received after executing action $\mathbf{a}_i$ in state $\mathbf{s}_i$. Specifically, we define the reward as the incremental increase in total generation delay incurred by assigning thought $i$:
\vspace{-.5mm}
\begin{align}\label{eq:DSACreward}
    r_i&= T^{\mathsf{actfin}}_{i}-T^{\mathsf{actfin}}_{i-1}\nonumber\\
    &-  \max\left(0,\ \frac{Score_{\min}}{|\mathcal{I}|} - Score_{i}^{\mathsf{actfin}}\right),
\end{align}
% \begin{align}\label{eq:DSACreward}
%     r_i&=\lambda\frac{\bar{T}^{\mathsf{loc}}-\Delta T_{tot}}{T^{\mathsf{loc}}}+(1-\lambda)\frac{\Delta Score_{tot}-\bar{Score}^{\mathsf{loc}}}{Score^{\mathsf{loc}}},\\
%     &\Delta T_{tot} = T_{tot\leftarrow \mathbf{X}_{1:i}}-T_{tot\leftarrow \mathbf{X}_{1:i-1}},\nonumber \\
%     &\Delta Score_{tot} = Score_{tot\leftarrow \mathbf{X}_{1:i}}-Score_{tot\leftarrow \mathbf{X}_{1:i-1}},\nonumber
% \end{align}
where $T^{\mathsf{actfin}}_{i-1}=0$ for the input thought $i=0$, $\frac{Score_{\min}}{|\mathcal{I}|}$ denotes the average quality threshold that each thought should attain, and $Score_{i}^{\mathsf{actfin}}$ denotes the actual generation quality of thought $i$ on its designated server. The penalty term becomes active when the quality of thought $i$ falls below the per-thought quality threshold, thereby implicitly enforcing constraint $\mathcal{C}2$ through penalization and effectively integrating it into the learning objective.

% $\bar{T}^{\mathsf{loc}}$ and $\bar{Score}^{\mathsf{loc}}$ denote the average generation delay and average generation quality for a thought, respectively, which are given by $\bar{T}^{\mathsf{loc}}=T^{\mathsf{loc}}/|\mathcal{I}|$ and $\bar{Score}^{\mathsf{loc}}=Score^{\mathsf{loc}}/|\mathcal{I}|$. Additionally, $T_{tot\leftarrow \mathbf{X}_{1:i}}$ and $Score_{tot\leftarrow \mathbf{X}_{1:i}}$ denote the generation delay and quality, respectively, given the thought assignment plan $\mathbf{X}_{1:i}=\{\mathbf{x}_{1}, \cdots, \mathbf{x}_i\}$.

\end{itemize}

\begin{figure}[t]
\includegraphics[width=.48\textwidth]{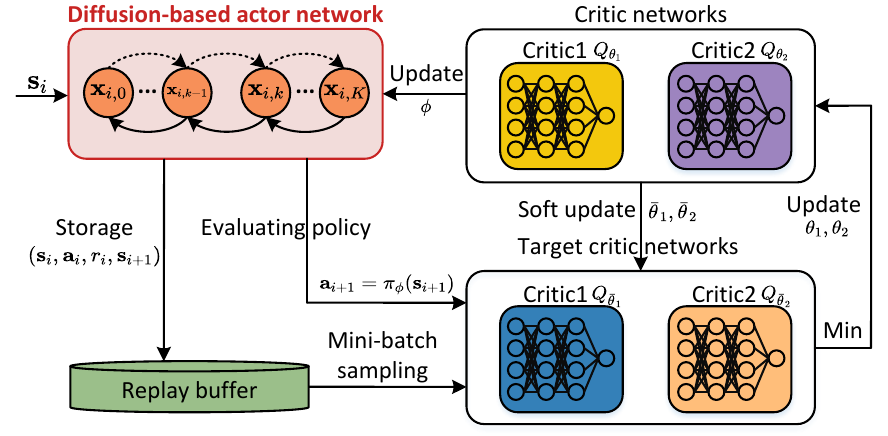}
\centering
\vspace{-1.5mm}
\caption{The overall architecture of the DSAC algorithm.}
\label{fig:DSAC}
\end{figure}

\subsection{Architecture of the DSAC Algorithm}  \label{subsec:DSACarchitecture}
The architecture of DSAC is illustrated in Fig.~\ref{fig:DSAC}. It comprises a diffusion model-based actor network, two critic networks, two target critic networks, and a replay buffer, as detailed below.

\begin{itemize}[leftmargin=4mm]
\item \textbf{\textit{Diffusion Model-Based Actor Network:}} In DSAC, the actor network $\pi_{\phi}$, parameterized by $\phi$, serves as the denoiser used in the reverse process of the diffusion model. Unlike the standard reinforcement learning objective, which focuses solely on maximizing the expected cumulative reward, DSAC further incorporates an entropy term. This enables the optimal policy to maximize its entropy, thereby maintaining sufficient exploration~\cite{haarnoja2018soft}:
\vspace{-.5mm}
\begin{align}\label{eq:optimalpolicy}
    \pi^*_{\phi}=\arg \max_{\pi_{\phi}}\sum_{i}\mathbb{E}_{(\mathbf{s}_i,\mathbf{a}_i)\sim{\pi_{\phi}}}[r_i+\varepsilon \mathcal{H}(\pi_{\phi}(\cdot | \mathbf{s}_i))],
\end{align}
where $\mathcal{H}(\pi_{\phi}(\cdot | \mathbf{s}_i))=\mathbb{E}_{\mathbf{a}_i\sim\pi_{\phi}}[-\log\pi_{\phi}(\mathbf{a}_i|\mathbf{s}_i)]$ denotes the entropy that measures the randomness of policy $\pi_{\phi}$, and $\varepsilon$ is the temperature parameter that controls the trade-off between the entropy term and the reward.

\item \textbf{\textit{Replay Buffer:}} During training, once thought $i$ is scheduled, the transition tuple $(\mathbf{s}_i, \mathbf{a}_i, r_i, \mathbf{s}_{i+1})$ is stored in the replay buffer $\mathcal{D}$. The replay buffer serves as an experience memory, enabling the agent to break temporal correlations through random mini-batch sampling, thereby improving sample efficiency and stabilizing the learning process.

\item \textbf{\textit{Two Critic Networks:}} Two critic networks, $Q_{\theta_1}$ and $Q_{\theta_2}$, parameterized by $\theta_1$ and $\theta_2$, respectively, take the state $\mathbf{s}_i$ and action $\mathbf{a}_i$ as inputs and produce the corresponding Q-values $Q_{\theta_n}(\mathbf{s}_i,\mathbf{a}_i)$~\cite{haarnoja2018soft}:
\vspace{-.5mm}
\begin{align}\label{eq:softactionvalue}
    Q_{\theta_n}(\mathbf{s}_i,\mathbf{a}_i)&=r_i+ \gamma\mathbb{E}_{\mathbf{s}_{i+1} \sim \mathcal{D}, \mathbf{a}_{i+1} \sim \pi_{\phi}}[Q_{\theta_n}(\mathbf{s}_{i+1},\mathbf{a}_{i+1}) \nonumber\\
    &-\varepsilon\log\pi_{\phi}(\mathbf{a}_{i+1}|\mathbf{s}_{i+1})], \ n\in \{1,2\}.
\end{align}
Specifically, $\gamma$ denotes the reward discount factor, and these Q-values quantify the expected quality of the state–action pair by incorporating both reward and entropy, where a higher value reflects not only a greater expected return but also a higher potential for exploration. Moreover, during policy improvement, the smaller of the two Q-values is used as the target Q-value, which helps reduce overestimation bias and stabilize training.

\item\textbf{\textit{Policy Improvement:}} After sufficient exploration, a mini-batch of samples is randomly drawn from the replay buffer $\mathcal{D}$ to update both the critic and actor networks. For the two critic networks, both are updated independently using the same optimization target. Specifically, the parameters $\theta_1$ and $\theta_2$ of the two critic networks are trained to minimize the soft Bellman residual~\cite{haarnoja2018soft}:
\vspace{-.5mm}
\begin{align}\label{eq:DSACcritic}
    J_{Q}(\theta_n)&=\mathbb{E}_{(\mathbf{s}_i, \mathbf{a}_i, r_i, \mathbf{s}_{i+1})\sim \mathcal{D}}\Big[\frac{1}{2}(Q_{\theta_n}(\mathbf{s}_i,\mathbf{a}_i) \nonumber\\ 
    &-\min_{n \in \{1,2\}} Q_{\bar{\theta}_n}(\mathbf{s}_i,\mathbf{a}_i))^2\Big], \ n \in \{1,2\},
\end{align}
where $\min_{n \in {1,2}} Q_{\bar{\theta}_n}(\mathbf{s}_i,\mathbf{a}_i)$ represents the soft Bellman target, which combines the immediate reward and the expected future return while incorporating an entropy bonus. It can be computed as~\cite{haarnoja2018soft}:
\vspace{-.5mm}
\begin{align}\label{eq:Bellmantarget}
    &\min_{n \in \{1,2\}} Q_{\bar{\theta}_n}(\mathbf{s}_i,\mathbf{a}_i)=r_i \nonumber\\
    &+\gamma \mathbb{E}_{ 
    \mathbf{a}_{i+1 \sim \pi_{\phi}}}\Big[
    \min_{n \in \{1,2\}} Q_{\bar{\theta}_n}(\mathbf{s}_{i+1},\mathbf{a}_{i+1}) 
    \nonumber\\
    &- \alpha\log\pi_{\phi}(\mathbf{a}_{i+1}|\mathbf{s}_{i+1})\Big],
\end{align}
where $Q_{\bar{\theta}_n}$ denotes the target critic networks, which take the state $\mathbf{s}_{i+1}$ and action $\mathbf{a}_{i+1}$ as inputs and produce the corresponding Q-values. Finally, the parameters $\phi$ of the actor network are learned by directly minimizing the expected Kullback–Leibler divergence~\cite{haarnoja2018soft}:
\vspace{-.5mm}
\begin{align}\label{eq:DSACactor}
   \hspace{-3mm} J_{\pi}(\phi) \!=\! \mathbb{E}_{\mathbf{s}_i \sim \mathcal{D}}\! \left[ 
    \mathbb{E}_{\mathbf{a}_i \sim \pi_{\phi}}\! 
    \left[ \alpha \log \pi_{\phi}(\mathbf{a}_i|\mathbf{s}_i) \!-\! Q_{\theta_1}(\mathbf{s}_i, \mathbf{a}_i) \right] 
\right].
\end{align}
To ensure stable training, the parameters of the target networks are updated gradually, allowing smooth changes in the learned Q-value estimates over time. This is achieved through soft updates as follows:
\vspace{-.5mm}
\begin{align}\label{eq:DSACtarget}
    \bar{\theta}_n\leftarrow \tau \theta_n+(1-\tau)\bar{\theta}_n, \ n \in \{1,2\}, 
\end{align}
where $\tau \in (0,1]$ denotes the update rate of the target networks.

\end{itemize}

\begin{algorithm} [!t]
\footnotesize
  \SetAlgoLined
  \SetKwData{Left}{left}\SetKwData{This}{this}\SetKwData{Up}{up}
  \SetKwFunction{Union}{Union}\SetKwFunction{FindCompress}{FindCompress}
  \SetKwInOut{Input}{input}\SetKwInOut{Output}{output}
  %\Input{A bitmap $Im$ of size $w\times l$}
  %\Output{A partition of the bitmap}
  \textbf{Input:} $\phi$, $\theta_1$, $\theta_2$, $\bar{\theta}_1 \leftarrow \theta_1$, $\bar{\theta}_2 \leftarrow \theta_2$, $\mathcal{D}\leftarrow \emptyset$.

  \textbf{Output:} The optimal thought assignment decisions.
  \BlankLine
  
  \For{$episode=1$ \KwTo $E$}{
        
        \For{$i=1$ \KwTo $|\mathcal{I}|$}{ 
            Obtain $\mathbf{s}_i$ according to~\eqref{eq:DSACstate} and initialize the assignment decision distribution $\mathbf{x}_{i,K} \sim \mathcal{N}(0, \mathbf{I})$.
      
            \For{$k=K$ \KwTo $0$} {
            Use a denoiser $\pi_{\phi}$ to infer the noise $\tilde{\mathbf{z}}_k=\pi_{\phi}(\mathbf{x}_{i,k},k,\mathbf{s}_i)$.
        
            Calculate the mean $\tilde{\mu}_{i,k}$ and the distribution $q(\mathbf{x}_{i,k-1}|\mathbf{x}_{i,k})$ by~\eqref{eq:reversemean} and~\eqref{eq:reversedistribution}, respectively. 

            Calculate the thought assignment decision distribution $\mathbf{x}_{i,k-1}$ by~\eqref{eq:reverseupdate}.
		  }
      
        Obtain the optimal assignment decision as $\mathbf{a}_i=\arg \max \{\mathbf{x}_{i,0}\}$.
        
        Receive the reward $r_i$ according to~\eqref{eq:DSACreward} and transition to the next state $\mathbf{s}_{i+1}$.
      
        Store the transition tuple $(\mathbf{s}_i, \mathbf{a}_i, r_i, \mathbf{s}_{i+1})$ into $\mathcal{D}$. 
        
        Randomly sample a batch of transitions, update the parameters $\theta_1$ and $\theta_2$ of the two critic networks using~\eqref{eq:DSACcritic}, update the parameter $\phi$ of the actor network using~\eqref{eq:DSACactor}, and update the parameters $\bar{\theta}_1$ and $\bar{\theta}_2$ of the two target critic networks using~\eqref{eq:DSACtarget}.
        }
    }
  \caption{DSAC Algorithm.}\label{algo:DSAC}
\end{algorithm}\DecMargin{1em}

\vspace{-3mm}
\subsection{DSAC Algorithm and Complexity Analysis}
\vspace{-.15mm}
Algorithm~\ref{algo:DSAC} provides the pseudocode of the proposed DSAC algorithm. The computational complexity of DSAC mainly stems from two aspects: training complexity and execution complexity. Note that since the training of learning-based methods can be carried out in a cloud data center with abundant computational resources~\cite{liu2024dnn,liu2026cross}, our analysis of the DSAC algorithm’s computational complexity primarily focuses on the execution phase.
 
During the execution phase, only the trained diffusion-based actor network is required, and the computational complexity corresponds to a forward pass through this network. Assume the denoiser in DSAC consists of $L$ fully connected layers, each with $N$ neurons. A single forward pass through the denoiser then requires $\mathcal{O}(LN^2)$ multiply–accumulate operations~\cite{du2024diffusion}. Since the diffusion model performs $K$ steps to generate the action for each thought, the per-thought complexity is $\mathcal{O}(KLN^2)$. Finally, across $E$ episodes, each containing $I$ thoughts to be scheduled, the total computational complexity of the proposed DSAC algorithm is $\mathcal{O}(EIKLN^2)$, which scales linearly with the number of episodes $E$, the number of thoughts $I$, the number of steps in the reverse process $K$, and quadratically with the width of the network’s hidden layers $N$.

\vspace{-3mm}
\section{Performance Evaluation}  \label{sec:simulation} 
\vspace{-.15mm}
In this section, we first present the simulation parameter settings and then evaluate the performance of the proposed DSAC by comparing it with three benchmark solutions.

\vspace{-3mm}
\subsection{Simulation Settings}
\vspace{-.15mm}

\vspace{-1.5mm}
\subsubsection{Network Layout} 
\vspace{-.15mm}

We consider a 100m $\times$ 100m square network, where the BS is located at the center and [0, 8] SPs are uniformly distributed over the area. The bandwidth allocated to the BS and each SP $B$ is 2MHz. The transmit power of the BS $p_0$ is set to 1W, while that of each SP $p_u$ is set to 0.1W~\cite{10301697}. The noise power spectral density $N_0$ is set to $4\times 10^{-21}$W/Hz. The output token count of each SP is modeled as a finite-state Markov process with three states, representing sufficient, moderate, and scarce computational availability, respectively. Specifically, the corresponding output token counts are set to 125, 100, 75, and 50 tokens. That is, at each time slot $t$, the available token count of  SP $u$, denoted by $C_{u,t}$, can take one of these three values. We assume that the transition probabilities among these three states are as follows:\footnote{The diagonal entries are the largest in each row, reflecting the tendency of SP computational availability to remain stable across consecutive time slots, while off-diagonal entries capture occasional transitions due to background workload fluctuations, with higher probabilities assigned to adjacent states to reflect the gradual nature of resource variation.}
\begin{align}
\text{Pr}^C =
\begin{bmatrix}
\text{Pr}^C_{11} & \text{Pr}^C_{12} & \text{Pr}^C_{13} & \text{Pr}^C_{14} \\
\text{Pr}^C_{21} & \text{Pr}^C_{22} & \text{Pr}^C_{23} & \text{Pr}^C_{24}\\
\text{Pr}^C_{31} & \text{Pr}^C_{32} & \text{Pr}^C_{33} & \text{Pr}^C_{34}\\
\text{Pr}^C_{41} & \text{Pr}^C_{42} & \text{Pr}^C_{43} & \text{Pr}^C_{44}\\
\end{bmatrix}
=
\begin{bmatrix}
0.4 & 0.3 & 0.2 & 0.1\\
0.3 & 0.4 & 0.2 & 0.1\\
0.1 & 0.2 & 0.4 & 0.3\\
0.1 & 0.2 & 0.3 & 0.4\\
\end{bmatrix}.
\end{align}
The main simulation parameters are summarized in Table~\ref{tab:simulatios}.

\vspace{-1.5mm}
\subsubsection{Algorithm Layout} 
\vspace{-.15mm}
We implement DSAC algorithm using Anaconda 25.11.0 with Python 3.12.0 and PyTorch 2.11.0 on a Windows platform equipped with an Intel Core i5-13600KF CPU. For the diffusion model, the denoiser is implemented as a multi-layer perceptron (MLP) comprising a time embedding module and a three-layer fully connected network. The time embedding module encodes the diffusion timestep via a sinusoidal positional embedding, projecting it into a 16-dimensional vector. This timestep embedding is then concatenated with the state and action inputs and fed through three fully connected layers with a hidden dimension of 400, ultimately producing the denoised action as output. 

% For all other hyperparameter settings of our algorithm, please refer to our open-source code at: \url{https://github.com/ZhangLiu/ToTAIGC}.

\begin{table}[!t]
\centering
\footnotesize
\caption{Parameters used in simulation~\cite{liu2023rfid,10024305,10301697}.}
\label{tab:simulatios}
\setlength{\tabcolsep}{6pt}
\begin{tabular}{ll}
\toprule
\textbf{Parameter} & \textbf{Value} \\
\midrule
Initial generation quality of the BS ($\sigma_0$) & 50\\
Rate of quality improvement of the BS ($\rho_0$) & 0.085\\
Generation delay per token of the BS ($\eta_0$) & 0.05\\
Initial generation overhead of the BS ($\psi_{0}$)& 0.1\\
Output token count of the BS ($C_{0}$)& 150\\
Initial generation quality of SP $u$ ($\sigma_u$) & (30, 55)\\
Rate of quality improvement of SP $u$ ($\rho_u$) & (0.035, 0.055)\\
Generation delay per token of SP $u$ ($\eta_u$) & (0.02, 0.04)\\
Initial generation overhead of SP $u$ ($\psi_{u}$)& (0.05, 0.15)\\
Date size associated with edge $e_{i,j}$ & [5, 10]KB\\
% Generation quality threshold $Score_{min}$ & $0.8 \times Score_{\mathsf{tot}}$\\
Number of episodes ($E$) & 1000\\
Reward discount factor ($\gamma$) & 0.99\\
Target network update rate ($\tau$) & 0.005\\
\bottomrule
\end{tabular}
% \vspace{-1.5mm}
\end{table}

\vspace{-3mm}
\subsection{Benchmark Solutions}
\vspace{-.15mm}
To demonstrate the effectiveness of the proposed D3PG algorithm, we have relied on three benchmark solutions:
\begin{itemize}[leftmargin=4.5mm] 
    \item \emph{SAC~\cite{haarnoja2018soft}:} We employ the soft actor-critic (SAC) algorithm to optimize thought assignment. As an off-policy actor-critic method, SAC improves exploration and training stability through entropy regularization. In contrast to our proposed method, SAC does not incorporate a diffusion model and instead relies on a conventional neural network policy to generate actions. This baseline is adopted to highlight the advantage of the diffusion-based decision mechanism.
    \item \emph{PPO~\cite{schulman2017proximal}:} We employ proximal policy optimization (PPO) to optimize thought assignment. As a representative state-of-the-art on-policy DRL algorithm, PPO updates its policy using trajectories collected from the current policy and stabilizes training via a clipped surrogate objective that constrains the policy update step. This baseline is included to benchmark our method against an advanced policy optimization approach.
    \item \emph{DDQN~\cite{van2016deep}:} We employ the double deep Q-network (DDQN) algorithm to optimize thought assignment. DDQN mitigates the overestimation bias in Q-learning by decoupling action selection from action evaluation through an online network and a target network, where the former selects the action and the latter evaluates its Q-value. This baseline is adopted to compare our method with a representative value-based learning scheme.
    % \item \emph{Local Generation (LG):} We consider a fully local generation scheme, where all thoughts are generated and evaluated exclusively at the BS. This approach leverages the strongest model parameters available at the BS, achieving the highest generation quality, but processes all thoughts sequentially without offloading to any SP, resulting in the largest generation delay. This baseline is included to reflect the quality upper bound and to highlight the latency cost of fully centralized generation.
\end{itemize}

% \begin{figure}[t!]
%     \centering
%     \includegraphics[width=.4\textwidth]{resultfigures/T2DRL_convergence_different_steps.pdf}
%     \vspace{-3mm}
%     \caption{Impact of actor learning rate on the reward in D3PG (number of V2V links $K=$, Lyapunov weight $V=$, and CSI feedback delay $T_{\text{delay}}=$).}
%     \label{fig:D3PG_convergency_different_LR}
%     \vspace{-3mm}
% \end{figure}

\begin{figure}[t!]
    \centering
    \includegraphics[width=.4\textwidth]{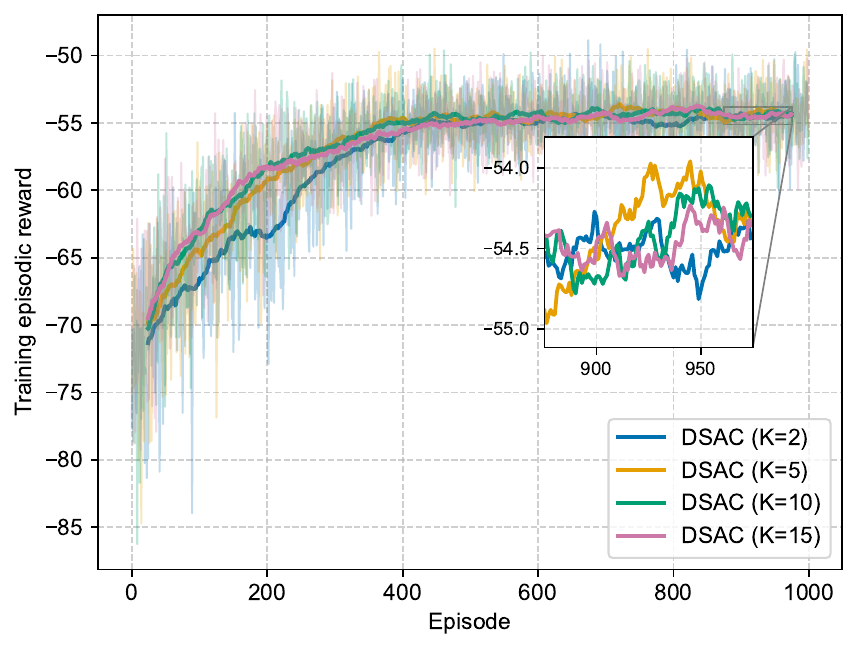}
    \vspace{-3mm}
    \caption{Impact of denoising step $K$ on the reward in DSAC (the number of SPs $U=6$, the number of ToT steps $ToT_{\mathsf{step}}=6$, and number of ToT thoughts per step $ToT_{\mathsf{thought}}=6$).}
    \label{fig:convergencydenoisingstep}
    \vspace{-3mm}
\end{figure}

\begin{figure}[t!]
    \centering
    \includegraphics[width=.4\textwidth]{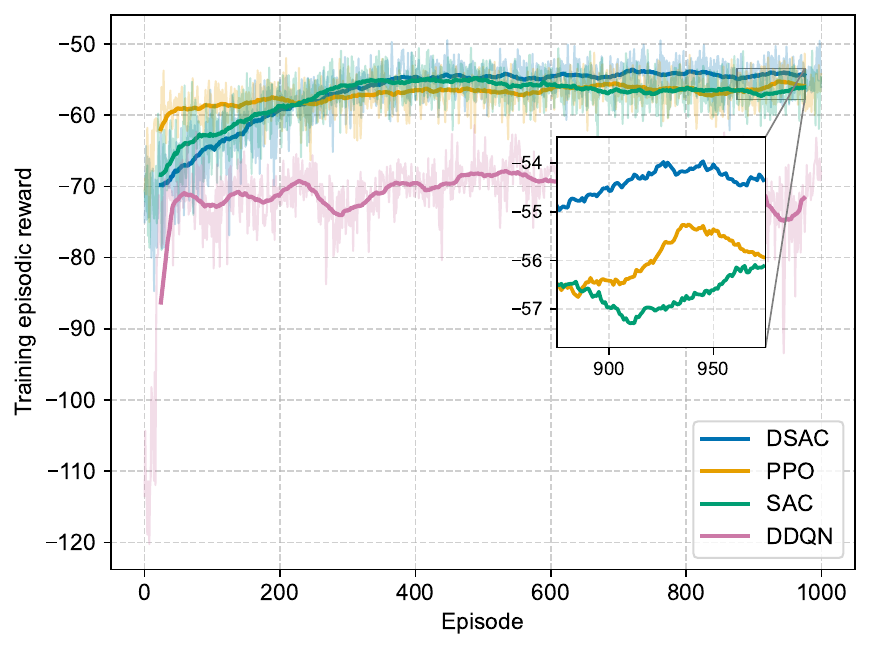}
    \vspace{-3mm}
    \caption{Comparison of reward curves among different algorithms (the number of SPs $U=6$, the number of ToT steps $ToT_{\mathsf{step}}=6$, and number of ToT thoughts per step $ToT_{\mathsf{thought}}=6$).}
    \label{fig:algoconvergency}
\end{figure}

\begin{figure*}[htbp]
    \centering
    \begin{subfigure}[b]{0.32\textwidth}  % 使用文本宽度的约32%
        \includegraphics[width=\linewidth]{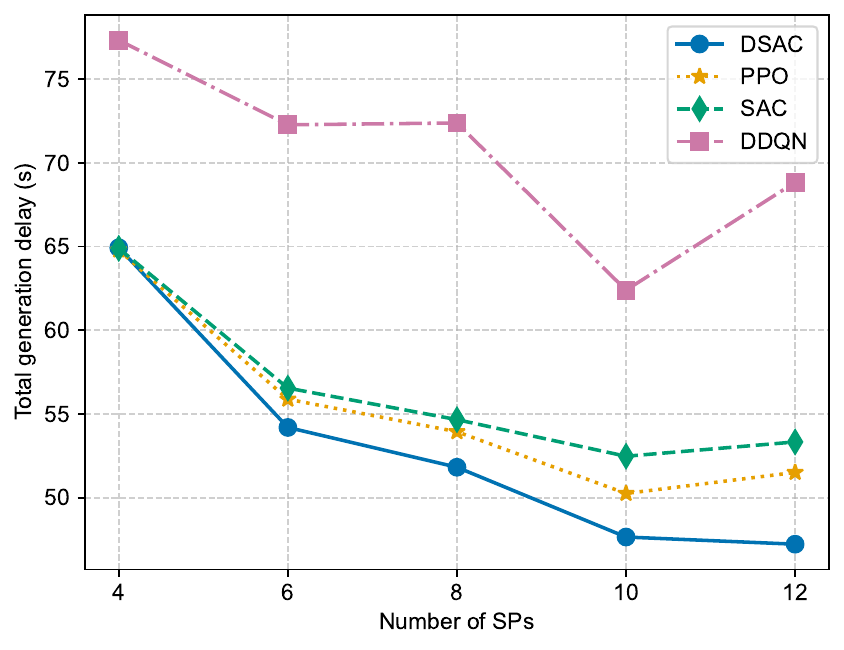}
        \vspace{-3.5mm}
        \caption{Total generation delay versus number of SPs (the number of ToT steps $ToT_{\mathsf{step}}=6$, and number of ToT thoughts per step $ToT_{\mathsf{thought}}=6$).}
        \label{fig:delay_vs_sps}
    \end{subfigure}
    \hfill
    \begin{subfigure}[b]{0.32\textwidth}  % 使用文本宽度的约32%
        \includegraphics[width=\linewidth]{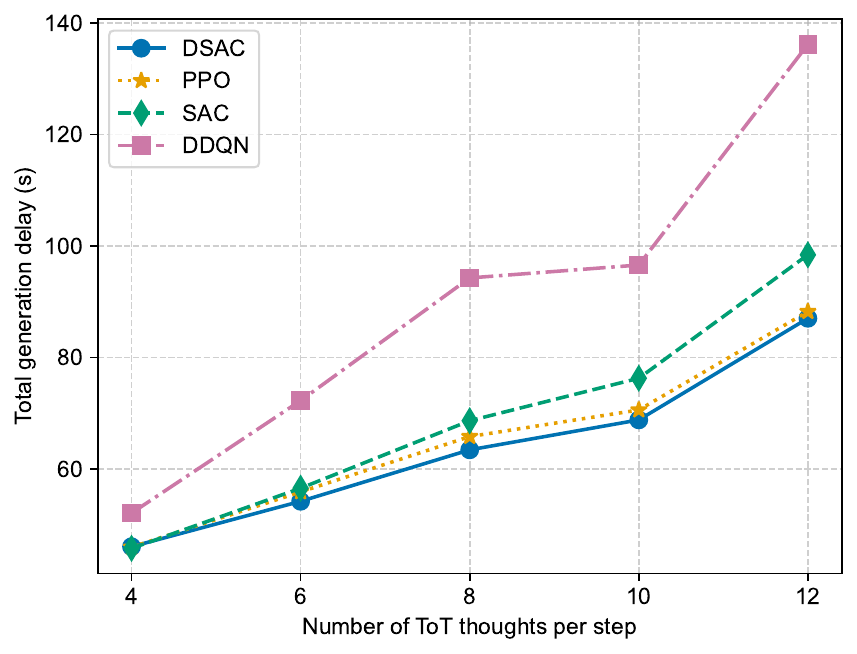}
         \vspace{-3.5mm}
        \caption{Total generation delay versus the number of ToT thoughts per step (the number of SPs $U=6$, and the number of ToT steps $ToT_{\mathsf{step}}=6$).}
        \label{fig:delay_vs_thoughts}
    \end{subfigure}
    \hfill
    \begin{subfigure}[b]{0.32\textwidth}  % 使用文本宽度的约32%
        \includegraphics[width=\linewidth]{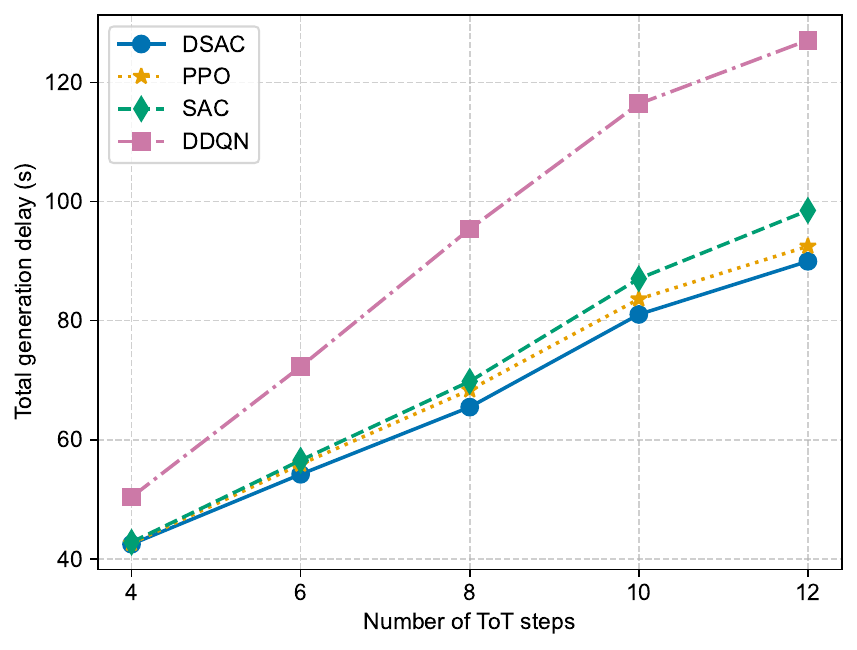}
         \vspace{-3.5mm}
        \caption{Total generation delay versus the number of ToT steps (the number of SPs $U=6$, and number of ToT thoughts per step $ToT_{\mathsf{thought}}=6$).}
        \label{fig:delay_vs_layers}
    \end{subfigure}
    \caption{Total generation delay under different simulation settings.}
    \label{fig:delay_performance}
    \vspace{-6.5mm}
\end{figure*}

\vspace{-3mm}
\subsection{Training Convergence Performance}
\vspace{-.15mm}
To eliminate the influence of randomness and ensure a fair comparison, we run each algorithm five times under different environmental settings (i.e., using five different random seeds) and use the average results to generate the following figures and tables.

% \subsubsection{Effect of the Value of Actor Learning Rate $\sigma^{\text{actor}}$} In Fig.~\ref{fig:D3PG_convergency_different_LR}, we illustrate the convergence behavior of the D3PG algorithm under different actor learning rates. The results show that the converged reward initially improves as the learning rate increases, but then declines beyond a certain threshold. This is because a moderate increase in the learning rate helps the D3PG agent escape local optima, whereas an excessively high learning rate may cause the actor network to overfit to inaccurate value estimates from the critic network, leading to reward fluctuations or degraded performance. Based on these observations, we set the actor network's learning rate in D3PG to $\sigma^{\text{actor}}=$ for comparisons with benchmark solutions in the subsequent experiments.

\vspace{-1.5mm}
\subsubsection{Effect of the Numbers of Denoising Steps $K$}
\vspace{-.15mm}

In Fig.~\ref{fig:convergencydenoisingstep}, we present the convergence behavior of the DSAC algorithm under varying numbers of denoising steps $K$ in the diffusion model, which directly influences the action sampling process. The results show that the converged reward improves as $K$ increases from 2 to 5, yet declines when $K$ is further increased to 10 and 15. This is because a moderate number of denoising steps stabilizes training and allows the diffusion model to capture more generalizable features. However, an excessive number of steps may over-smooth the output, removing useful signal components and ultimately degrading performance. Furthermore, the denoising step count $K$ has a significant impact on convergence speed: as $K$ increases, each action sample requires more iterative denoising steps, which raises the computational cost per episode and slows down the overall training process, ultimately leading to slower convergence. Therefore, we set $K=5$ in the following experiments to balance convergence speed and overall performance of DSAC.

\vspace{-1.5mm}
\subsubsection{Training Reward for Different Algorithms}
\vspace{-.15mm}
In Fig.~\ref{fig:algoconvergency}, we depict the convergence behavior of four different algorithms as the number of training episodes increases. The results show that DSAC achieves the highest episodic reward among all methods, with a notably more stable training curve. This superiority stems from the adoption of a diffusion-based actor network, which replaces the conventional MLP used in SAC. Unlike MLP-based actors that generate actions through a single forward pass -- often suffering from limited exploration capability and susceptibility to local optima in complex environments -- the diffusion-based actor generates actions through an iterative denoising process. This allows for progressive refinement and stochastic exploration of the action space, enabling the policy to more effectively navigate complex solution landscapes, avoid premature convergence, and ultimately converge to higher-quality actions with greater training stability.

PPO and SAC both perform worse than DSAC. The performance gap between SAC and DSAC can be primarily attributed to SAC's reliance on an MLP-based actor, which lacks the expressive power and exploratory capacity of a diffusion-based policy, leading to suboptimal action generation in complex environments. PPO, on the other hand, exhibits a smoother and more stable training curve than SAC, owing to its clipped surrogate objective that constrains the magnitude of policy updates at each training step and prevents destabilizing gradient steps. SAC, by contrast, continuously updates its policy based on stochastic gradient estimates without such explicit constraints, making it more susceptible to fluctuations during training.

DDQN exhibits the worst overall performance among all compared methods, characterized by the lowest converged reward and severe oscillations throughout training. This is because DDQN lacks an explicit policy representation and instead derives actions indirectly by selecting the argmax over estimated Q-values, making it less capable of capturing fine-grained behavioral patterns in complex environments. Furthermore, bootstrapped Q-value updates introduce training instability -- particularly when the reward landscape is non-stationary or the state space is high-dimensional -- resulting in persistent oscillations in the reward curve and an inability to converge to competitive reward levels.

\vspace{-1.5mm}
\subsubsection{Effect of the Number of SPs}
\vspace{-.15mm}
Fig.~\ref{fig:delay_vs_sps} illustrates the impact of the number of SPs on the total generation delay. As the number of SPs grows, all methods exhibit a general downward trend, since more SPs provide greater offloading opportunities, enabling increased parallelism in thought generation and thereby reducing overall latency. It is also worth noting that when the number of SPs increases from 10 to 12, PPO, SAC, and DDQN exhibit a slight rise in generation delay. This is because, beyond a certain point, the marginal benefit of additional offloading opportunities diminishes, while the increased coordination overhead and communication cost among more SPs begin to outweigh the parallelism gains. In contrast, DSAC maintains a consistently decreasing trend, as its diffusion-based policy is better able to navigate this trade-off and identify a more efficient assignment plan under the expanded SP set. Specifically, when the number of SPs is set to 6, DSAC achieves delay reductions of 3.02\% over PPO, 4.15\% over SAC, and 25.03\% over DDQN, and these margins further widen to 8.32\%, 11.45\%, and 31.39\%, respectively, when the number of SPs increases to 12.

\vspace{-1.5mm}
\subsubsection{Effect of the Number of ToT Thoughts per Step}
\vspace{-.15mm}
Fig.~\ref{fig:delay_vs_thoughts} illustrates the impact of the number of ToT thoughts per step on the total generation delay. As the number of thoughts per step increases, all methods exhibit a consistent upward trend in generation delay, since a larger number of thoughts per step directly increases the total number of model invocations required, thereby imposing a heavier computational and communication burden on the system. Notably, DDQN experiences a significantly steeper increase compared to the other methods, as its value-based discrete action mechanism struggles to efficiently allocate a growing number of thoughts across SPs, leading to increasingly suboptimal assignment decisions. In contrast, DSAC consistently achieves the lowest delay across all configurations. Specifically, when the number of thoughts per step is set to 6, DSAC achieves delay reductions of 3.02\% over PPO, 4.15\% over SAC, and 25.03\% over DDQN, and these margins further widen to 1.36\%, 11.57\%, and 36.09\%, respectively, as the number of thoughts per step increases to 12.

\vspace{-1.5mm}
\subsubsection{Effect of the Number of ToT Steps}
\vspace{-.15mm}
Fig.~\ref{fig:delay_vs_layers} illustrates the impact of the number of ToT steps on the total generation delay. As the number of ToT steps increases, all methods exhibit a consistent and approximately linear upward trend in generation delay, since each additional ToT step introduces a new layer of thoughts that must be generated and evaluated, directly amplifying the total number of model invocations and thus the overall latency. DDQN again stands out with a markedly steeper growth rate, reflecting its inherent limitations in handling the increasingly complex sequential assignment decisions introduced by deeper ToT structures. In contrast, DSAC, PPO, and SAC maintain comparatively moderate growth rates, with DSAC consistently achieving the lowest delay across all configurations. Specifically, when the number of ToT steps is set to 4, DSAC achieves delay reductions of 0.32\% over PPO, 0.77\% over SAC, and 15.71\% over DDQN, and these margins further widen to 2.73\%, 8.67\%, and 29.21\%, respectively, as the number of ToT steps increases to 12.

% \begin{table}[htbp]
% \centering
% \footnotesize
% \caption{Comparison of algorithm running time per time slot (millisecond).}
% % \vspace{-2mm}
% \label{table2}
% \rowcolors{1}{lightblue!30}{white}
% \begin{tabular}{|c|c|c|c|c|c|}
% \hline
% \textbf{Number of Users} & \textbf{10} & \textbf{12} & \textbf{14} & \textbf{16} & \textbf{18} \\ \hline\hline
% T2DRL & 1.112 & 1.162 & 1.197 & 1.226 & 1.272\\ \hline
% DDPG-based T2DRL & 0.302 & 0.317 & 0.324 & 0.339 & 0.351 \\ \hline
% SCHRS & 624.9 & 769.7 & 844.1 & 973.2 & 1128.7 \\ \hline
% \end{tabular}
% \end{table}
\begin{figure}[t!]
    \centering
    \includegraphics[width=.41\textwidth]{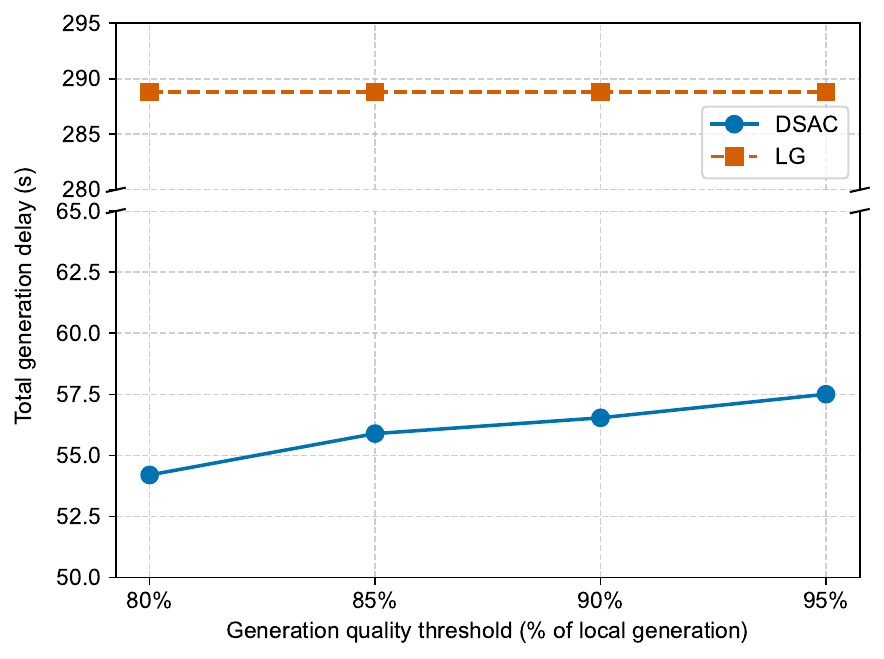}
    \vspace{-3mm}
    \caption{Total generation delay versus Generation Quality Threshold $Score_{min}$ (the number of SPs $U=6$, the number of ToT steps $ToT_{\mathsf{step}}=6$, and number of ToT thoughts per step $ToT_{\mathsf{thought}}=6$).}
    \label{fig:quality_threshold}
\end{figure}

\vspace{-1.5mm}
\subsubsection{Effect of Generation Quality Threshold $Score_{min}$} 
\vspace{-.15mm}
Fig.~\ref{fig:quality_threshold} presents the total generation delay of DSAC under varying generation quality thresholds, alongside the local generation (LG) baseline, where all thoughts are generated and evaluated exclusively at the BS. As the quality threshold increases from 80\% to 95\% of local generation quality,\footnote{Since generating all thoughts locally at the BS -- which hosts the most capable GenAI model -- yields the highest generation quality, the user can set the quality threshold as a desired percentage of this local generation quality (e.g., 80\% to 95\%) without requiring any prior quality measurements. This percentage-based formulation is both intuitive and flexible.} the generation delay of DSAC rises gradually. This is expected, since a higher quality threshold forces the system to assign more thoughts to the BS, which hosts the strongest generative model, thereby reducing the degree of offloading and limiting the parallelism gains. Despite this upward trend, DSAC maintains a substantially lower generation delay compared to LG across all quality thresholds. This is because LG processes all thoughts sequentially at the BS without any offloading, resulting in a consistently high generation delay regardless of the quality requirement. In contrast, DSAC achieves a delay reduction of 81.23\% over LG at the 80\% threshold, and still maintains a reduction of 80.09\% at the 95\% threshold. These results demonstrate that the proposed ToT thought assignment mechanism effectively navigates the trade-off between generation quality and latency, delivering significant delay savings even under stringent quality requirements.

\begin{table}[htbp]
\centering
\footnotesize
\caption{Comparison of algorithm running time per thought (milliseconds).}
\label{algo_runningtime}
\setlength{\tabcolsep}{6pt}
\begin{tabular}{cccccc}
\toprule
\textbf{Number of SPs} & \textbf{4} & \textbf{6} & \textbf{8} & \textbf{10} & \textbf{12} \\
\midrule
DSAC  & 5.39 & 6.12 & 6.45 & 5.81 & 4.62 \\
SAC  & 0.71 & 0.78 & 0.79 & 0.67 & 0.75 \\
PPO  & 1.28 & 1.51 & 1.48 & 1.41 & 1.27 \\
DDQN & 0.24 & 0.22 & 0.23 & 0.22 & 0.24 \\
\bottomrule
\end{tabular}
\end{table}

\vspace{-1.5mm}
\subsubsection{Algorithm Running Time Performance} 
\vspace{-.15mm}
Table~\ref{algo_runningtime} presents the running time per thought of each algorithm under varying numbers of SPs. DSAC incurs the highest running time among all methods, primarily due to its diffusion-based reverse process, which generates actions through a step-wise denoising procedure that introduces additional computation per thought assignment decision. PPO follows with the second highest running time, attributable to its on-policy nature, which requires repeated trajectory collection and policy updates. SAC and DDQN exhibit the lowest running times owing to their relatively lightweight network structures. Notably, the running time of DSAC does not increase monotonically with the number of SPs, suggesting that the diffusion-based policy can handle larger-scale SP configurations without a proportional rise in computational cost, and thus scales well to more complex deployment scenarios. Nevertheless, given that DSAC consistently achieves the lowest generation delay across all configurations subject to a user-adjustable quality constraint, we conclude that it offers superior performance with only a modest increase in computational overhead.

\vspace{-3mm}
\section{Conclusion} \label{sec:conclusion}
\vspace{-.15mm}
In this paper, we have investigated edge-enabled AIGC service provisioning with ToT prompting. Specifically, we have first characterized the number of output tokens as a measure of computational resources in GenAI models and established its mathematical relationship with generation delay and quality through extensive experiments with Qwen 2.5-7B-Instruct. Building on this foundation, we have introduced a DAG model to accurately capture the multi-step, dependency-aware reasoning process of ToT prompting, and formulated the DAG-based thought assignment problem as an integer nonlinear programming problem aimed at minimizing generation delay subject to a user-adjustable quality constraint. To solve this problem effectively, we have proposed the DSAC algorithm, which innovatively integrates diffusion models into the SAC framework to enable iterative refinement and stochastic exploration of the action space, thereby improving overall decision quality. Through extensive simulations, we have demonstrated that DSAC consistently outperforms all benchmark solutions across various simulation settings.

There are several limitations and future research directions in this work. First, this work considers a single job owner with one AIGC task. An interesting yet challenging extension is to support multiple job owners with concurrent AIGC tasks, where inter-task resource contention and scheduling fairness introduce additional complexity. Second, this work focuses on creative writing as a representative AIGC task. Extending the proposed framework to other content types, such as image and audio generation, by identifying appropriate computational resource proxies and fitting the corresponding mathematical relationships, represents a promising direction.
\vspace{-3mm}

% Future research could further explore the potential of utilizing an expert dataset which can be obtained offline through the brute-force search method, to conduct the forward process of diffusion models. This data set includes the optimal solutions for DNN partitioning and task offloading. Subsequently, a form of supervised learning could be applied to train the diffusion model to fit the action distribution generated by the reverse process and the expert data. Furthermore, investigations of competitions and cooperations among multiple RSUs for task acquisition is an enticing direction. 
% \vspace{-3mm}

\bibliographystyle{IEEEtran}
\bibliography{mainbody.bbl}
\vspace{-14mm}

\end{document}